\pdfoutput=1
\documentclass[journal]{IEEEtran}

\usepackage{microtype}
\usepackage{amsmath,amsfonts,amssymb}
\usepackage{amsthm}
\usepackage{mathtools}
\usepackage{graphicx}
\usepackage{booktabs}
\usepackage{array}
\usepackage{cite}
\usepackage{braket}
\usepackage{algorithm}
\usepackage[noend]{algpseudocode}
\usepackage{xcolor}
\usepackage[hidelinks]{hyperref}
\usepackage{orcidlink}

\newcommand{\bibdoi}[1]{\href{https://doi.org/#1}{\nolinkurl{#1}}}
\newcommand{\E}{\mathbb{E}}

\newcommand{\1}{\mathbf{1}}
\newcommand{\cD}{\mathcal{D}}

\newcommand{\vmax}{v^\star}
\newcommand{\vrob}{v^{\mathrm{rob}}}
\newcommand{\pw}{p^{w}}
\newcommand{\Dmax}{D_{\max}}
\newcommand{\Dadd}{D_{\mathrm{add}}}
\newcommand{\wmax}{w_{\max}}
\newcommand{\wmin}{w_{\min}}
\DeclareMathOperator{\Tr}{Tr}
\DeclareMathOperator{\supp}{supp}
\DeclareMathOperator{\STAB}{STAB}

\theoremstyle{plain}
\newtheorem{theorem}{Theorem}
\newtheorem{proposition}{Proposition}
\newtheorem{lemma}{Lemma}
\newtheorem{corollary}{Corollary}
\theoremstyle{definition}
\newtheorem{definition}{Definition}
\newtheorem{hypothesis}{Hypothesis}
\theoremstyle{remark}
\newtheorem{remark}{Remark}

\title{Where to Hide the Traps: Robust Placement in Verifiable Blind
Quantum Computation}

\author{
    \IEEEauthorblockN{
        Noureldin Mohamed\IEEEauthorrefmark{1}\,\orcidlink{0009-0001-4150-8690} and
        Saif Al-Kuwari\IEEEauthorrefmark{1}\,\orcidlink{0000-0002-4402-7710}
    }
    \vspace{0.2cm} \\
    \IEEEauthorblockA{
        \IEEEauthorrefmark{1}\textit{Qatar Center for Quantum Computing, College of Science and Engineering,} \\
        \textit{Hamad Bin Khalifa University, Doha, Qatar.} \\
        Corresponding author: nomo89098@hbku.edu.qa
    }
}

\begin{document}

\maketitle

\begin{abstract}
Trap-based verification of blind quantum computation hides test qubits
among the positions of a measurement pattern, and existing protocols choose
the trap configuration without regard to the delegated computation. We
model trap
placement as a zero-sum game between a client, who chooses a trap
distribution, and a server, who chooses where to apply a Pauli deviation,
with a damage profile $w$ and a per-trap detection constant $c$. Since the
client cannot know how damage from several positions combines, we solve the
game for all monotone subadditive damage functions consistent with $w$ at
once. In a single round, placing traps in proportion to $w$ guarantees
soundness error $1-c/\|w\|_1$ and is optimal up to a factor $1+\max_iw_i$,
while the optimistic solution can fail completely. Over $T$
test rounds, the error decays exponentially against a server that repeats
the same deviation, but a general server can keep it at
$\Theta(\|w\|_1/(cT))$, and against such servers the damage-weighted
placement is exactly optimal. Damage profiles follow from the flow of
the pattern: for brickwork patterns with $m=4h+1$ columns, only $(2h+1)^2$
positions can affect a decision qubit, whatever the width. With $k$ traps
per round, placing them on this set reduces the number of test rounds by the
same factor as for a single trap; with unrestricted trap sets and flat
damage we recover the known optimality of fractional coloring. Finally,
trap statistics certify the aggregate damage with a number of rounds
independent of the number of positions $N$, whereas locating the damage
requires $\Omega(N\epsilon^{-2}\log\delta^{-1})$ rounds.
\end{abstract}


\section{Introduction}
\label{sec:intro}

\subsection{Delegated quantum computation and its two security goals}

Quantum processors of useful size are expensive, few in number and
operated centrally. As a result, most users will run quantum algorithms
on hardware they neither own nor control. Two security requirements follow.
The first is \emph{privacy}: the server should learn nothing about the
algorithm, its input or its output. The second is \emph{integrity}: the
client should be able to decide whether the returned result is correct.
Privacy was addressed first. Childs showed that a client able to store
qubits and apply Pauli operations can hide a delegated
computation~\cite{childs_secure_2005}, and Broadbent, Fitzsimons and
Kashefi introduced universal blind quantum computation
(UBQC)~\cite{broadbent_universal_2009}, in which a client that can prepare
single-qubit states delegates an arbitrary computation in the
measurement-based model while revealing only an upper bound on its size. A
photonic implementation followed~\cite{barz_demonstration_2012}, and the
protocol family is surveyed in Ref.~\cite{fitzsimons_private_2017}.

\subsection{Integrity is the harder requirement}

Blindness limits what the server can learn, but it does not prevent the
server from deviating. A malicious or faulty server may deviate arbitrarily, and for computations that
are hard to simulate classically the client cannot recompute the answer.
Verification of quantum computation has become a field of its own,
with several approaches~\cite{gheorghiu_verification_2019}. Interactive
proofs with a verifier that holds a constant number of qubits show that
verification is possible in principle~\cite{aharonov_interactive_2017}.
Protocols with two non-communicating provers allow a fully classical
verifier to test entangled devices~\cite{reichardt_classical_2013}. Under
post-quantum computational assumptions a single quantum prover can be
verified by a classical client~\cite{mahadev_classical_2018}. Post-hoc
protocols verify a completed computation by estimating the energy of a
local Hamiltonian~\cite{fitzsimons_post_2018}, and other schemes certify the
resource state itself~\cite{hayashi_verifiable_2015,takeuchi_resource_2019}.
One approach combines verification with blindness, requires only
single-qubit preparation from the client, and has been implemented on
hardware~\cite{barz_experimental_2013,drmota_verifiable_2024}. This
approach, trap-based verification, is the subject of this paper.

\subsection{Trap-based verification and the placement question}

Trap-based verification uses a second consequence of blindness: the server
cannot tell which qubits carry the computation. The client hides
\emph{trap} qubits, whose measurement outcomes it can predict, among the
computation qubits. A
deviation risks disturbing a trap, and the client aborts when a trap
returns an unexpected value~\cite{fitzsimons_unconditionally_2017}.
Fitzsimons and Kashefi (FK) placed one trap uniformly at random among $N$
qubits and proved that the resulting protocol is
$(1-\tfrac{1}{2N})$-verifiable in general and $(1-\tfrac1N)$-verifiable for
classical output~\cite[Theorem~8]{fitzsimons_unconditionally_2017}, then
amplified soundness with fault-tolerant encodings. Later work reduced the
resource overhead~\cite{kashefi_optimised_2017}, established robustness and
device-independence properties~\cite{gheorghiu_robustness_2015}, and
obtained composable, noise-robust protocols whose only overhead is
repetition by interleaving test rounds with computation
rounds~\cite{leichtle_verifying_2021}. Trappified schemes provide a general
formalism for such constructions and relate verification to quantum error
detection~\cite{kapourniotis_unifying_2024}. The same machinery now supports
verified delegation of variational
algorithms~\cite{inajetovic_verifiable_2025}, and round counts are being
optimized against security targets~\cite{vandam_optimizing_2026}.

All of these protocols sample a trap configuration from a distribution,
and the distributions used in
Refs.~\cite{fitzsimons_unconditionally_2017,leichtle_verifying_2021,kapourniotis_unifying_2024}
are uniform. FK sample the trap position uniformly. Leichtle \emph{et al.}
sample a color class of a proper coloring uniformly. The
trappified-scheme formalism allows a general distribution over trap
configurations~\cite{kapourniotis_unifying_2024}, and Kapourniotis
\emph{et al.}\ already optimized it for standard traps: the largest
worst-case detection probability of a test round is $1/\chi_f(G)$, the
inverse fractional chromatic number of the resource graph, and an optimal
fractional coloring attains it, which improves on uniform coloring whenever
$\chi_f(G)<\chi(G)$~\cite[Sec.~6.2.1]{kapourniotis_unifying_2024}. That
optimization treats every position as equally important to protect.
Uniform or coloring-based sampling is the right choice when every
position is equally important to protect, but in measurement patterns this
is rarely the case. A deviation on a qubit whose outcome
feeds a wide cone of corrections can corrupt most of the output, whereas a
deviation on a qubit outside the causal past of the qubit carrying the
answer cannot affect it at all. To our knowledge, no previous work has asked
how the placement distribution should depend on this structure, how to
make that dependence robust to the unknown way in which damage combines, or
how much is gained by optimizing it, in particular when the number of
traps per round is limited.

\subsection{Hypotheses}
\label{sec:hypotheses}

The paper is organized around three working hypotheses, which the later
sections test.

\begin{itemize}
\item \textbf{W1 (structured damage).} For realistic measurement patterns the
  damage caused by a single-qubit deviation varies strongly across
  positions, and the profile of these damages is computable by the client
  in polynomial time from public data.
\item \textbf{W2 (placement leverage).} Sampling traps in proportion to damage
  rather than uniformly improves the guaranteed soundness error by a factor
  that grows with the non-uniformity of the profile, and the improvement is
  close to the best achievable by any distribution.
\item \textbf{W3 (model robustness).} Guarantees that depend on a particular rule
  for combining damage across positions are unsafe. Only guarantees that
  hold uniformly over a family of combination rules can be certified by a
  client that does not know the rule.
\end{itemize}

Section~\ref{sec:flow} supports W1: damage profiles follow from the flow of
the pattern, and for brickwork patterns only $((m+1)/2)^2$ of the $nm$
positions can affect a decision qubit, and this fraction shrinks as the
pattern gets wider. Sections~\ref{sec:robust}
and~\ref{sec:budget} support W2, with a guarantee in which $N$ is replaced
by $\|w\|_1$, a converse within a factor $1+\wmax$ for a single round, and
exact optimality of the damage-weighted placement against general servers.
Section~\ref{sec:game} supports W3 with a placement that is optimal under one
combination rule and has soundness error one under another. These
working hypotheses are distinct from the modeling hypotheses H1--H5 of
Section~\ref{sec:model}, which formalize the properties of trap-based
protocols that the analysis assumes.

\subsection{Placement as a zero-sum game}

The client chooses a randomized placement, the server chooses where to
deviate, neither observes the other's choice, and the gain of one is the
loss of the other. This is a two-player zero-sum
game~\cite{vonneumann_theory_1944} with the structure of a \emph{security
game}~\cite{tambe_security_2011,korzhyk_stackelberg_2011}, in which a
defender randomizes scarce resources over targets of unequal value and an
attacker strikes where protection is weakest relative to value. Traps
correspond to monitoring resources, qubits to targets, and damage to target
value.

A Pauli deviation acts on a set of positions at once, as an attacker with
several resources does in the security games of
Ref.~\cite{korzhyk_multiple_2011}. What is specific to our setting is that
the client does not know how damage combines over the attacked set.
Assuming that a deviation on a set $S$ causes the damage of its worst
position reduces the server to single-qubit attacks and yields a closed-form
solution. We show in Section~\ref{sec:model} that this rule is the most
optimistic one consistent with the single-qubit data. Its optimal placement
leaves low-damage positions untrapped, and a server that deviates on all of
those positions simultaneously corrupts the output without any risk of
detection. The analysis must therefore hold uniformly over a family of
rules, in the spirit of robust security games with payoff
uncertainty~\cite{kiekintveld_interval_2013}.

\subsection{Contributions}
\label{sec:contrib}

The main contribution is robust, computation-dependent trap placement:
choosing where to put traps from the damage structure of the delegated
pattern, with guarantees that hold whatever rule combines damage across
positions. All improvements reported below are measured against
computation-independent placement \emph{with the same number of traps per
round}. For single-trap rounds that baseline is uniform placement; for rounds
with $k$ traps it is the corresponding computation-independent construction
of Proposition~\ref{prop:ktraps}. When each test round may use a full color
class, as in Ref.~\cite{leichtle_verifying_2021}, the computation-independent
optimum is the fractional-coloring result of
Ref.~\cite{kapourniotis_unifying_2024}, which we recover as
Theorem~\ref{thm:fractional}, and for the brickwork decision profiles studied
here it leaves no room for improvement (Corollary~\ref{cor:brick}).

The results fall into four groups.

\paragraph{Model and single-round game (Sections~\ref{sec:model}--\ref{sec:robust})}
In the hidden test-round template, and under hypotheses that formalize
standard properties of trap-based protocols, the soundness error is bounded
by the payoff of a placement game (Lemma~\ref{lem:reduction}); for flat
damage the per-round bound is $1-c/N$, which coincides numerically with both
FK constants. For every monotone subadditive damage function consistent
with the profile $w$, the value lies between an optimistic value $\vmax$
and a robust value $\vrob$ (Proposition~\ref{prop:sandwich}). The
optimistic problem is an elementary linear program with a water-filling
solution (Theorem~\ref{thm:minimax}), and its placement fails against
servers that accumulate damage (Proposition~\ref{prop:optimism}). The
damage-weighted placement $\pw=w/\|w\|_1$ certifies $1-c/\|w\|_1$ for every
damage function in the family, and no placement improves its detection
margin by more than $1+\wmax\le2$ (Theorem~\ref{thm:robust}). The robust
value has an exact dual in terms of damage loads
(Theorem~\ref{thm:duality}), a closed form for two-level profiles
(Proposition~\ref{prop:twolevel}), and can be computed for hundreds of
positions by column generation with a knapsack oracle
(Section~\ref{sec:robust-alg}).

\paragraph{Repetition (Section~\ref{sec:budget})} Against persistent
deviations the soundness error decays exponentially in the number $T$ of
test rounds, and the two envelopes coincide once $\vmax(T)\le\wmin/2$
(Theorem~\ref{thm:collapse}). Against general servers the error is at most
$\rho_D(p)/(c(T+1))$, where $\rho_D(p)=\max_SD(S)/p(S)$, and for single-position
attacks this is tight up to a factor four. The minimum of $\rho_D$ equals $\|w\|_1$ for every
$D\in\cD(w)$ and is attained only by $\pw$ (Theorem~\ref{thm:general}), so
against a malicious server the damage-weighted placement is exactly optimal,
whatever the combination rule.

\paragraph{Damage from structure (Sections~\ref{sec:flow}
and~\ref{sec:trapsets})} Flow cones computed from the flow give sound damage
functions (Proposition~\ref{prop:influence}),
and in exact state-vector simulations of blind brickwork patterns they
coincide with the set of positions whose deviation changes the output
(Section~\ref{sec:flow-sim}). For brickwork patterns with $m=4h+1$ columns, the
flow cone of a decision qubit away from the boundary rows has exactly
$(2h+1)^2=((m+1)/2)^2$ positions, independently of the width $n$
(Proposition~\ref{prop:cone}). The number of test rounds therefore falls by
a factor close to $N/|B|=4nm/(m+1)^2$, about $17$ for $G_{65\times13}$ in
both regimes, relative to uniform single-trap placement.
With at most $k$ traps per round the same factor holds for every $k$ up
to half the cone (Proposition~\ref{prop:ktraps}); for $G_{65\times13}$,
$24$ traps placed on the cone nearly match the per-round guarantee that a
full color class of about $420$ traps gives. For unrestricted trap sets and flat
damage, Theorem~\ref{thm:fractional} recovers the fractional-coloring
optimum of Ref.~\cite{kapourniotis_unifying_2024} in our notation, and
Corollary~\ref{cor:brick} shows that uniform coloring is then robustly
optimal for brickwork decision profiles.

\paragraph{Statistical certification and protocol (Sections~\ref{sec:tomography}
and~\ref{sec:protocol})} Trap flags estimate flag rates that bracket the
marginals of the twirled deviation. The damage-weighted placement is the
unique range-optimal design for certifying aggregate damage, with sample
complexity independent of $N$, whereas every strategy, adaptive or not,
needs $\Omega(N\epsilon^{-2}\log\delta^{-1})$ rounds to learn all flag rates
(Theorem~\ref{thm:lower}). The components combine into a protocol with
explicit soundness statements for both regimes (Theorem~\ref{thm:protocol}),
valid when the client aborts on any flagged trap. The noise-tolerant
threshold rule is a statistical certification procedure; with a single
computation round it is not sound in the sense of
Section~\ref{sec:model-template} (Proposition~\ref{prop:notsound}).

We checked every closed-form statement against exact linear programs over
all $2^N-1$ supports or against exact simulation.

\subsection{Organization}

Section~\ref{sec:prelim} reviews measurement-based computation and flow,
blind computation, trap-based verification, Pauli twirling, zero-sum and
security games, and the statistical and combinatorial tools used later, and
positions the work against prior results. Section~\ref{sec:model} states the
modeling hypotheses and the damage formalism.
Sections~\ref{sec:game}--\ref{sec:budget} solve the placement game.
Section~\ref{sec:flow} computes damage profiles from the flow,
Section~\ref{sec:trapsets} treats rounds with many traps, and
Section~\ref{sec:tomography} develops statistical certification from
trap flags.
Section~\ref{sec:tension} compares detection with reconstruction,
Section~\ref{sec:protocol} assembles the protocol, and
Sections~\ref{sec:discussion} and~\ref{sec:conclusion} discuss limitations
and conclude.

\section{Preliminaries and Related Work}
\label{sec:prelim}

This section collects the background used in the rest of the paper and
positions our contribution. Readers familiar with blind verification can
skip to Section~\ref{sec:prelim-position}.

\subsection{Notation}
\label{sec:prelim-notation}

We write $[N]=\{1,\dots,N\}$ and $\Delta_N$ for the probability simplex on
$[N]$. For $p\in\Delta_N$ and $S\subseteq[N]$, $p(S)=\sum_{i\in S}p_i$. For a
vector $w\in[0,1]^N$, $w(S)=\sum_{i\in S}w_i$, $\|w\|_1=w([N])$,
$\wmax=\max_iw_i$ and $\wmin=\min_iw_i$. $X,Y,Z$ are the single-qubit Pauli
operators, and for an $n$-qubit Pauli operator $E$, $\supp E$ is the set of
qubits on which it acts non-trivially. $CZ_{uv}$ is the controlled-$Z$ gate,
and $\1[\cdot]$ is the indicator function. For probability measures $P,Q$,
$\mathrm{KL}(P\|Q)$ is the Kullback--Leibler divergence, and
$\mathrm{kl}(a\|b)$ is its value for Bernoulli distributions with means
$a,b$. For a graph $G=(V,E)$, $\mathcal N(u)$ is the neighborhood of vertex $u$.
Table~\ref{tab:notation} summarizes the main symbols used in the paper and
points to where each is defined.

\begin{table}[!t]
\centering
\caption{Main notation and where each symbol is defined.}
\label{tab:notation}
\small
\setlength{\tabcolsep}{4pt}
\begin{tabular}{@{}>{\raggedright\arraybackslash}p{0.18\columnwidth}>{\raggedright\arraybackslash}p{0.74\columnwidth}@{}}
\toprule
Symbol & Meaning\\
\midrule
$N$, $[N]$ & number of positions; index set\\
$p\in\Delta_N$ & trap placement distribution; $p(S)=\sum_{i\in S}p_i$\\
$c$ & per-trap detection constant (Hypothesis~\ref{ass:detect})\\
$E$, $S$ & Pauli deviation and its support\\
$w$, $\|w\|_1$ & damage profile and its effective size\\
$\wmax$, $\wmin$ & largest and smallest entries of $w$\\
$D$, $\cD(w)$ & damage function; class of monotone subadditive functions consistent with $w$\\
$\Dmax,\Dadd$ & optimistic and conservative extremes, \eqref{eq:extremal}\\
$U_D(p,S)$, $U^{(T)}_D$ & single-round and persistent payoffs, \eqref{eq:payoff}, \eqref{eq:payoffT}\\
$\bar D_S$ & shorthand for $\Dadd(S)$ in Section~\ref{sec:robust-dual}\\
$\vmax$, $\vrob$ & optimistic and robust game values\\
$p^\star$, $\pw$ & water-filling and damage-weighted placements\\
$\kappa(w)$ & fractional packing number of covers\\
$L_i(\sigma)$ & damage load of position $i$, \eqref{eq:loads}\\
$T$, $\kappa$ & number of test rounds; index of the computation round\\
$\rho_D(p)$ & damage-to-coverage ratio, \eqref{eq:rho}\\
$\mathcal N(u)$ & graph neighborhood of $u$\\
$f$, $F(u)$ & flow; flow cone of $u$, \eqref{eq:flowdamage}\\
$B$, $h$ & flow cone of the decision qubit; $m=4h+1$\\
$q_i$, $r_i$ & deviation marginal and flag rate at position $i$\\
$\xi_t$, $Y_t$, $\beta_p$ & flag of test round $t$; importance-weighted statistic and its range\\
$\chi_f(G)$ & fractional chromatic number\\
\bottomrule
\end{tabular}
\end{table}

\subsection{Measurement-based quantum computation and flow}
\label{sec:prelim-mbqc}

In the one-way model~\cite{raussendorf_oneway_2001,briegel_mbqc_2009} a
computation is driven by single-qubit measurements on an entangled resource.
Given a graph $G=(V,E)$, the \emph{graph state}~\cite{hein_graph_2006} is
\begin{equation}
  \ket{G}=\prod_{(u,v)\in E}CZ_{uv}\,\ket{+}^{\otimes V},
  \label{eq:graphstate}
\end{equation}
and the state \eqref{eq:graphstate} is stabilized by $K_u=X_u\prod_{v\in\mathcal N(u)}Z_v$ for every $u\in V$.
Computations use measurements in the $X$--$Y$ plane of the Bloch sphere,
with basis $\{\ket{+_\alpha},\ket{-_\alpha}\}$ where
$\ket{\pm_\alpha}=(\ket0\pm e^{i\alpha}\ket1)/\sqrt2$. The outcome is
$s\in\{0,1\}$.

The elementary step is one-bit teleportation. Measuring the first qubit of
$CZ_{12}(\ket\psi\otimes\ket+)$ at angle $\alpha$ with outcome $s$ leaves the
second qubit in $X^sHP(-\alpha)\ket\psi$, where $P(\beta)=\mathrm{diag}(1,e^{i\beta})$
and $H$ is the Hadamard gate. Measurements therefore apply gates, up to
Pauli \emph{byproduct} operators that depend on random outcomes. Later
measurements compensate for these byproducts. An $X$ byproduct is absorbed
by flipping the sign of a later measurement angle, and a $Z$ byproduct by
adding $\pi$ to it.

A \emph{measurement pattern} is specified by an \emph{open graph}
$(G,I,O)$, with input and output vertex sets $I,O\subseteq V$, together with
angles $\{\phi_u\}_{u\in O^c}$ for the measured qubits. Whether the
byproducts can always be corrected is captured by
flow~\cite{danos_determinism_2006}.

\begin{definition}[Flow~\cite{danos_determinism_2006}]
\label{def:flow}
A \emph{flow} on $(G,I,O)$ is a map $f:O^c\to I^c$ with a partial order
$\prec$ on $V$ such that, for every $u\in O^c$: (i) $f(u)\in\mathcal N(u)$;
(ii) $u\prec f(u)$; and (iii) $u\prec k$ for every $k\in\mathcal N(f(u))\setminus\{u\}$.
\end{definition}

If $(G,I,O)$ has a flow, measuring in any order compatible with $\prec$ and
adapting each angle as
\begin{equation}
  \phi'_j=(-1)^{s_{f^{-1}(j)}}\phi_j+\pi\!\!\!\bigoplus_{k:\,j\in\mathcal N(f(k))\setminus\{k\}}\!\!\! s_k
  \label{eq:adapt}
\end{equation}
makes the pattern deterministic, that is, independent of the measurement
outcomes~\cite{danos_determinism_2006,danos_measurement_2007}. Here the
first term is omitted when $j$ has no flow predecessor. Operationally, the
outcome $s_u$ influences only $f(u)$, through the sign, and
$\mathcal N(f(u))\setminus\{u\}$, through the $\pi$ shift. Condition~(iii)
guarantees that all of these qubits are measured after $u$. Generalized
flow (gflow) relaxes Definition~\ref{def:flow} and characterizes a broader
class of deterministic patterns~\cite{browne_generalized_2007}. The influence
structure defined by \eqref{eq:adapt} is the basis of the damage profiles in
Section~\ref{sec:flow}.

\paragraph{Brickwork states} The resource used in UBQC is the
\emph{brickwork state} $G_{n\times m}$ with $n$ rows and $m\equiv5\pmod8$
columns~\cite{broadbent_universal_2009}. All qubits in the same row are
joined in a line. For each column $j\equiv3\pmod8$ and each odd row $i$,
vertical edges join $(i,j)$ to $(i+1,j)$ and $(i,j+2)$ to $(i+1,j+2)$. For
each column $j\equiv7\pmod8$ and each even row $i$, the same two vertical
edges are added. The row map $f(i,j)=(i,j+1)$, with columns ordered left
to right, is a flow. With angles in $\{k\pi/4\}_{k=0}^7$, brickwork patterns
are universal for quantum computation~\cite{broadbent_universal_2009}. The
graph is bipartite, since the parity of $i+j$ is a proper $2$-coloring; we
use this in Section~\ref{sec:trapsets}.

\subsection{Universal blind quantum computation}
\label{sec:prelim-ubqc}

UBQC~\cite{broadbent_universal_2009} proceeds as follows for a brickwork
pattern with angles $\{\phi_u\}$.
\begin{enumerate}
\item The client prepares every qubit $u$ in $\ket{+_{\theta_u}}$ with
      $\theta_u$ uniform in $\{k\pi/4\}$ and sends the qubits to the server.
\item The server applies $CZ$ along every edge of $G_{n\times m}$.
\item In flow order, the client computes $\phi'_u$ from
      \eqref{eq:adapt}, samples $r_u\in\{0,1\}$ uniformly, and sends
      $\delta_u=\phi'_u+\theta_u+r_u\pi$. The server measures qubit $u$ in
      the basis $\{\ket{\pm_{\delta_u}}\}$ and returns $b_u$. The client sets
      $s_u=b_u\oplus r_u$.
\end{enumerate}
The angle $\theta_u$ hides $\phi'_u$, and $r_u$ hides the outcome. The
server's view, consisting of the graph dimensions and the transcript of
angles and outcomes, is independent of the computation and of the input,
so the protocol is \emph{blind}. It leaks only $(n,m)$. Blindness of UBQC is
composable in the abstract-cryptography framework~\cite{dunjko_composable_2014}.
Variants include a measurement-only client that receives qubits from the
server~\cite{morimae_blind_2013}. Blindness was first demonstrated with
photons~\cite{barz_demonstration_2012}. A review is given in
Ref.~\cite{fitzsimons_private_2017}.

\subsection{Trap-based verification}
\label{sec:prelim-traps}

\paragraph{Soundness} Let the client either output a result or abort. A
delegation protocol is \emph{$\varepsilon$-verifiable}~\cite{fitzsimons_unconditionally_2017}
if, for every server strategy, the probability that the client accepts an
output that differs from the ideal one is at most $\varepsilon$. In FK this
is written $\Tr(P_{\mathrm{incorrect}}\rho_{\mathrm{out}})\le\varepsilon$,
where $P_{\mathrm{incorrect}}$ projects onto incorrect output states
together with the accepting trap outcomes. Throughout the paper we call
$\varepsilon$ the \emph{soundness error}, and $1-\varepsilon$ the
\emph{detection margin} when it is guaranteed against the worst deviation.

\paragraph{Traps and dummies} A \emph{trap} is a qubit prepared in
$\ket{+_\theta}$, all of whose graph neighbors are \emph{dummies} prepared
in computational basis states $\ket{d}$ with random $d\in\{0,1\}$. Since
$CZ(\ket{+_\theta}\otimes\ket d)=Z^d\ket{+_\theta}\otimes\ket d$, entangling
leaves the trap in the product state $\ket{+_{\theta+\pi\sum d}}$, which the
client knows. The client asks for a measurement at angle
$\theta+\pi\sum d+r\pi$, and the honest outcome is $b=r$. Blindness makes
traps, dummies and computation qubits indistinguishable to the server.

\paragraph{Single-graph constructions} FK embed the computation together
with traps in an enlarged graph, with the trap position sampled uniformly,
and prove the single-trap bounds quoted in
Section~\ref{sec:intro}~\cite[Theorem~8]{fitzsimons_unconditionally_2017}.
Exponentially small soundness error then comes from encoding the
computation, so that an undetected corruption requires many simultaneous
errors, each of which risks hitting a trap. Kashefi and Wallden reduce the
overhead of this construction~\cite{kashefi_optimised_2017}. Gheorghiu,
Kashefi and Wallden establish robustness and device-independence
properties~\cite{gheorghiu_robustness_2015}.

\paragraph{Test rounds} Leichtle, Music, Kashefi and Ollivier (LMKO) take a
different route~\cite{leichtle_verifying_2021}. The client runs many rounds
on the \emph{same} graph. Some are \emph{computation rounds}, and the rest,
chosen at random, are \emph{test rounds} that the server cannot distinguish
from computation rounds. In a test round the client fixes a proper coloring
of the graph, chooses a color uniformly at random, places traps on every
vertex of that color and dummies everywhere else. The client aborts if the
number of failed test rounds exceeds a threshold, set by the tolerated noise
level, and otherwise outputs the majority result of the computation rounds.
For bounded-error decision problems this yields composable security with
exponentially small soundness error, tolerance to a bounded level of noise,
and no overhead beyond repetition~\cite{leichtle_verifying_2021}. The
number of rounds, the fraction of test rounds and the threshold can be
tuned for a target security level~\cite{vandam_optimizing_2026}. The
template of Section~\ref{sec:model-template}, with one computation round and
abort on any failed test, is the simplest member of this family; the
majority-vote version is discussed in Section~\ref{sec:discussion}.

\paragraph{Trappified schemes} Kapourniotis, Kashefi, Leichtle, Music and
Ollivier abstract these constructions~\cite{kapourniotis_unifying_2024}. A
\emph{trappified canvas} is a partial pattern together with a trap, meaning
a test on part of the graph. A trappified scheme is a set of canvases with
a common order, an embedding of the computation, and a probability
distribution over the canvases. The scheme \emph{$\epsilon$-detects} an
error set $\mathcal E$ if every $\mathsf e\in\mathcal E$ triggers a trap with
probability at least $1-\epsilon$ over the choice of canvas. This notion
links verification to error detection. The formalism allows a general
distribution over canvases~\cite{kapourniotis_unifying_2024}, and
Section~6.2.1 of Ref.~\cite{kapourniotis_unifying_2024} optimizes it for
standard traps placed on independent sets: the best worst-case detection
probability is $1/\chi_f(G)$, attained by an optimal fractional coloring,
with the five-cycle as an example where this beats every proper coloring.
This optimization is independent of the embedded computation. How the
distribution should depend on the computation, and how to make it robust to
an unknown damage combination rule, is not addressed.

\paragraph{Experiments and applications} Verifiable blind computation has
been demonstrated with photons~\cite{barz_experimental_2013} and with a
trapped-ion server networked to a photonic client~\cite{drmota_verifiable_2024}.
Related verification of measurement-based sampling has been performed with
trapped ions~\cite{ringbauer_verifiable_2025}. Verifiable delegated
variational algorithms interleave test rounds across
iterations~\cite{inajetovic_verifiable_2025}. Resource-cost optimization of
noise-robust protocols tunes round-level parameters for target
security~\cite{vandam_optimizing_2026}.

\subsection{Pauli twirling}
\label{sec:prelim-twirl}

Any channel $\mathcal E$ on $n$ qubits can be expanded in the Pauli basis as
$\mathcal E(\rho)=\sum_{P,Q}\chi_{PQ}\,P\rho Q^\dagger$. Averaging over
conjugation by a uniformly random Pauli operator $R$ removes the
off-diagonal terms:
\begin{equation}
  \frac1{4^n}\sum_R R^\dagger\,\mathcal E(R\rho R^\dagger)\,R=\sum_P\chi_{PP}\,P\rho P .
  \label{eq:twirl}
\end{equation}
The right-hand side of \eqref{eq:twirl} is a \emph{Pauli channel}, a probabilistic mixture of Pauli
errors~\cite{dankert_exact_2009,nielsen_quantum_2010}. Twirling is the
mechanism behind randomized benchmarking and randomized
compiling~\cite{wallman_noise_2016}. In trap-based verification the
client's secret parameters, $\theta$, $r$ and the dummy bits, play the role
of the twirl. This is how the soundness analyses of
Refs.~\cite{fitzsimons_unconditionally_2017,leichtle_verifying_2021} reduce an
arbitrary server deviation to a mixture of Pauli attacks. We use this
reduction as Hypothesis~\ref{ass:twirl}. It also explains why the
estimation questions of Section~\ref{sec:tomography} concern Pauli error
rates, for which efficient learning methods
exist~\cite{flammia_efficient_2020,harper_efficient_2020,harper_fast_2021}
and whose limits are characterized~\cite{chen_learnability_2023}.

\subsection{Zero-sum games and security games}
\label{sec:prelim-games}

In a finite two-player zero-sum game the row player chooses a mixed strategy
$x\in\Delta_m$, the column player chooses $y\in\Delta_n$, and the column
player receives $x^\top My$. The minimax
theorem~\cite{vonneumann_theory_1944} states that
\begin{equation}
  \min_{x}\max_{y}x^\top My=\max_{y}\min_{x}x^\top My ,
  \label{eq:minimax}
\end{equation}
and Sion's theorem extends \eqref{eq:minimax} to quasi-convex--quasi-concave payoffs on
compact convex sets~\cite{sion_general_1958}. The common value and the
optimal strategies can be computed by linear programming. A pair of
optimal strategies is a Nash equilibrium. If one player commits publicly
first (a \emph{Stackelberg} game), its guaranteed value in a zero-sum game
equals the minimax value, so commitment neither helps nor hurts.

\emph{Security games}~\cite{tambe_security_2011} are the defender--attacker
models behind deployed systems for airport checkpoints, air-marshal
scheduling and coast-guard patrols. A defender allocates limited resources
at random over targets of different value, and an attacker observes the
allocation and strikes a target. The relation between the defender's
Stackelberg and Nash strategies was characterized by Korzhyk
\emph{et al.}~\cite{korzhyk_stackelberg_2011}. Closely related formulations
place sensors in networked control systems under
attack~\cite{nguyen_zerosum_2022}. Two features of our setting have
counterparts in this literature. Attackers that hit several targets at once
are studied by Korzhyk, Conitzer and Parr~\cite{korzhyk_multiple_2011},
with utilities that add over the attacked targets; they compute Nash
equilibria in polynomial time and show that optimal Stackelberg commitment
becomes NP-hard, so that the two solution concepts no longer coincide.
Uncertainty in the attacker's
payoffs, handled by optimizing against the worst case in an interval, is
studied by Kiekintveld, Islam and Kreinovich~\cite{kiekintveld_interval_2013}.
In our setting the uncertainty is not in the value of each target but in the
rule that combines values across the attacked set, and we handle it with the
monotone subadditive envelope of Section~\ref{sec:model-damage}. The optimal
solutions under the optimistic rule have the \emph{water-filling} form
familiar from power allocation over parallel Gaussian
channels~\cite{cover_elements_2006}.

Game theory appears in delegated quantum computation in two other ways. In
\emph{rational} delegation, the server is paid according to a reward
function under which honesty is its best strategy, so no traps are
needed~\cite{morimae_rational_2020,takeuchi_sumcheck_2019}.
\emph{Quantum games} study players who use quantum
strategies~\cite{meyer_quantum_1999,eisert_quantum_1999}. Our server is
adversarial and unbounded, and both players randomize classically. Quantum
mechanics enters only through the payoff.

\subsection{Statistical and combinatorial tools}
\label{sec:prelim-tools}

\paragraph{Concentration and lower bounds} For independent $Y_t\in[a,b]$,
Hoeffding's inequality gives
$\Pr[|\bar Y-\E\bar Y|\ge\epsilon]\le2\exp(-2T\epsilon^2/(b-a)^2)$~\cite{hoeffding_probability_1963}.
The Azuma--Hoeffding inequality extends this to martingale differences with
predictable bounded ranges~\cite{azuma_weighted_1967}. For lower bounds we
use the Bretagnolle--Huber inequality, $P(A)+Q(A^c)\ge\tfrac12e^{-\mathrm{KL}(P\|Q)}$
for every event $A$~\cite[Lemma~2.6]{tsybakov_introduction_2009}, together
with the divergence decomposition for adaptively sampled
observations~\cite[Lemma~15.1]{lattimore_bandit_2020}. Certifying a scalar
property of a quantum device is typically much cheaper than full
reconstruction~\cite{flammia_direct_2011,eisert_certification_2020}. We
make this separation precise for traps.

\paragraph{Independent sets and colorings} An \emph{independent set} of
$G$ contains no edge. The \emph{stable-set polytope} $\STAB(G)$ is the
convex hull of indicator vectors of independent sets. The \emph{fractional
chromatic number}
\begin{equation}
  \chi_f(G)=\min\Bigl\{\textstyle\sum_I y_I:\ \sum_{I\ni v}y_I\ge1\ \forall v,\ y\ge0\Bigr\}
  \label{eq:chif}
\end{equation}
satisfies $\omega(G)\le\chi_f(G)\le\chi(G)$, where $\omega$ is the clique
number and $\chi$ the chromatic number~\cite{scheinerman_fractional_1997}.
For perfect graphs, including bipartite graphs, all three coincide, and
linear optimization over $\STAB(G)$ is polynomial~\cite{grotschel_geometric_1988}.
For bipartite graphs,
$\STAB(G)=\{\pi\ge0:\pi_u+\pi_v\le1\ \forall uv\in E\}$~\cite{schrijver_combinatorial_2003}.
The minimum knapsack problem, which appears later as a server best response,
is NP-hard~\cite{garey_computers_1979}.

\subsection{Positioning of this work}
\label{sec:prelim-position}

Table~\ref{tab:prior} summarizes how trap configurations are sampled in the
works closest to ours. Existing analyses fix the sampling distribution and
bound soundness for it. We instead optimize the distribution itself, using
the structure of the computation, in a way that is robust to the unknown
rule by which damage combines. For flat damage and unrestricted trap
sets our formulation reduces to the fractional-coloring problem solved in
Ref.~\cite{kapourniotis_unifying_2024}; the new ingredients are the damage
profile, the robustness to the combination rule, the multi-round analysis,
and rounds with a limited number of traps. The results apply within each
framework in Table~\ref{tab:prior}. Sections~\ref{sec:game}--\ref{sec:budget}
cover instances with a bounded number of traps, and
Section~\ref{sec:trapsets} covers rounds with many traps. The work is
complementary to round-level optimization~\cite{vandam_optimizing_2026}.
Other verification paradigms~\cite{aharonov_interactive_2017,reichardt_classical_2013,mahadev_classical_2018,fitzsimons_post_2018,hayashi_verifiable_2015,takeuchi_resource_2019}
do not tie a test to a position, so they do not face this placement
decision. On the estimation side, our contribution is specific to traps. We
identify the flag rates as the quantities a trap round estimates, show that
they bracket but do not identify the channel marginals, and prove matching
upper and lower bounds that separate certification from localization.

\begin{table}[t]
\centering
\caption{How trap configurations are sampled in prior trap-based
verification and in this work.}
\label{tab:prior}
\small
\renewcommand{\arraystretch}{1.15}\setlength{\tabcolsep}{3pt}
\begin{tabular}{@{}>{\raggedright\arraybackslash}p{0.2\columnwidth}>{\raggedright\arraybackslash}p{0.37\columnwidth}>{\raggedright\arraybackslash}p{0.37\columnwidth}@{}}
\toprule
Work & Trap configuration & Sampling distribution\\
\midrule
FK~\cite{fitzsimons_unconditionally_2017} & single trap in an enlarged graph & uniform over positions\\
LMKO~\cite{leichtle_verifying_2021} & color class of a proper coloring, per test round & uniform over colors\\
Trappified schemes~\cite{kapourniotis_unifying_2024} & set of trappified canvases & general distribution allowed; fractional coloring optimal for standard traps~\cite[Sec.~6.2.1]{kapourniotis_unifying_2024}; not tied to the computation\\
van Dam \emph{et al.}~\cite{vandam_optimizing_2026} & round-level parameters & within-round design taken as given\\
\textbf{This work} & positions or independent sets, any number of traps per round & optimized against damage profile; robust over damage models\\
\bottomrule
\end{tabular}
\end{table}

\section{Model}
\label{sec:model}

\subsection{Protocol template}
\label{sec:model-template}

We work in the hidden test-round setting of
Refs.~\cite{leichtle_verifying_2021,kapourniotis_unifying_2024}. In this
setting the damage profile of the computation and the trap distribution are
defined on the same set of qubits. Let $G$ be the resource
graph of the delegated pattern and let its vertices be the
\emph{positions} $[N]$. The client runs $T+1$ rounds on $G$. One round,
chosen uniformly at random and kept secret, is the \emph{computation
round}, in which the pattern is executed blindly. The other $T$ rounds are
\emph{test rounds}. In a test round the client draws a position $t\sim p$
from a \emph{placement} $p\in\Delta_N$, prepares a trap at $t$ and dummies at
every other position, and learns whether the trap \emph{flags}. The client
aborts if any test round flags and otherwise outputs the result of the
computation round. Test rounds with many traps are treated in
Section~\ref{sec:trapsets}; protocols with several computation rounds and a
majority vote are discussed in Section~\ref{sec:discussion}.

\paragraph{Correctness event} Let $O^\star$ be the register that carries
the answer. Following Ref.~\cite{fitzsimons_unconditionally_2017}, the
output of the computation round is \emph{incorrect} when the state of
$O^\star$ has a component orthogonal to the ideal output state, that is, the
event is defined by the projector $\1-\ket{\psi_{\mathrm{ideal}}}\bra{\psi_{\mathrm{ideal}}}$
on $O^\star$ together with acceptance. This event concerns the state
produced by the protocol and not the bounded-error answer of the underlying
algorithm, so an honest server never produces it. A protocol is
\emph{$\varepsilon$-sound} if, for every server strategy, the probability of
accepting an incorrect output is at most $\varepsilon$; this is the notion
of $\varepsilon$-verifiability of Ref.~\cite{fitzsimons_unconditionally_2017}.

\subsection{Modeling hypotheses}
\label{sec:model-hyp}

The following hypotheses formalize the properties of trap-based protocols
used in the analysis. Hypotheses~\ref{ass:twirl}--\ref{ass:damage} are
established in, or abstract, the soundness analyses of
Refs.~\cite{fitzsimons_unconditionally_2017,leichtle_verifying_2021,kapourniotis_unifying_2024}.
Hypothesis~\ref{ass:rounds} is a restriction on the server that we use only
where stated. Section~\ref{sec:model-status} discusses their status.

\begin{hypothesis}[Pauli reduction]
\label{ass:twirl}
In each round, the client's one-time padding makes the server's deviation
equivalent, from the client's point of view, to an honest execution followed
by a probabilistic mixture of Pauli operators $E$ applied to the qubits
before measurement~\cite{fitzsimons_unconditionally_2017,leichtle_verifying_2021}.
We write $S=\supp E\subseteq[N]$.
\end{hypothesis}

\begin{hypothesis}[Blindness of the round structure]
\label{ass:blind}
The server's view is independent of which round is the computation round
and of the trap positions, for every placement $p$, and the client reveals
no flag before the end of the protocol. Hence the sequence of deviations
$(E_1,\dots,E_{T+1})$ is independent of the computation-round index and of
the trap positions.
\end{hypothesis}

\begin{hypothesis}[Local detection]
\label{ass:detect}
There is a constant $c\in(0,1]$ such that, in a test round with trap
position $t$ and deviation $E$, the trap flags with probability at least $c$
if $t\in\supp E$ and does not flag if $t\notin\supp E$. Flag events of
different traps, in the same or different rounds, are independent given the
deviations.
\end{hypothesis}

\begin{hypothesis}[Damage bound]
\label{ass:damage}
There is a set function $D:2^{[N]}\to[0,1]$ with $D(\emptyset)=0$ such that
a computation round with deviation $E$ produces an incorrect output with
probability at most $D(\supp E)$, independently of the flag events of the
test rounds given the deviations.
\end{hypothesis}

\begin{hypothesis}[Persistent deviation]
\label{ass:rounds}
The deviations of the different rounds have a common support: the server
draws a support $S$ once, possibly at random, and deviates on $S$ in every
round.
\end{hypothesis}

The independence statements in Hypotheses~\ref{ass:detect}
and~\ref{ass:damage} reflect the fact that each round, and within a round
each trap, uses fresh client secrets: the flag of a trap is governed by its
own parameters $(\theta_t,r_t)$, and the correctness of the computation round
by the secrets of the computation qubits.

\begin{lemma}[Persistent deviations]
\label{lem:reduction}
Under Hypotheses~\ref{ass:twirl}--\ref{ass:rounds}, the protocol with $T$ test
rounds satisfies
\begin{align}
  \Pr[\text{accept}\wedge\text{incorrect}]
  &\le\max_{S\neq\emptyset}U^{(T)}_D(p,S),\nonumber\\
  U^{(T)}_D(p,S)&:=D(S)\bigl(1-c\,p(S)\bigr)^T,
  \label{eq:payoffT}
\end{align}
where $p(S)=\sum_{i\in S}p_i$. For $T=1$ we write $U_D:=U^{(1)}_D$, so that
\begin{equation}
  U_D(p,S)=D(S)\bigl(1-c\,p(S)\bigr).
  \label{eq:payoff}
\end{equation}
\end{lemma}

\begin{proof}
By Hypothesis~\ref{ass:rounds} and linearity it suffices to bound the
payoff for a fixed support $S$ used in all rounds. The computation round
produces an
incorrect output with probability at most $D(S)$
(Hypothesis~\ref{ass:damage}). Each test round draws $t\sim p$ independently
of the deviation (Hypothesis~\ref{ass:blind}) and fails to flag with
probability at most $1-c\,p(S)$ (Hypothesis~\ref{ass:detect}). Independence
of these events gives $D(S)(1-c\,p(S))^T$.
\end{proof}

The single-round payoff \eqref{eq:payoff} is the object of
Sections~\ref{sec:game} and~\ref{sec:robust}. It describes the per-round
trade-off between the damage a deviation causes when it hits the
computation round and the probability that it is caught when it hits a test
round. Section~\ref{sec:budget} treats $T>1$, first under
Hypothesis~\ref{ass:rounds} and then for general servers, for which the
relevant functional changes from \eqref{eq:payoffT} to the damage-to-coverage
ratio $\max_SD(S)/p(S)$.

\begin{remark}[Calibration of $c$]
\label{rem:calibration}
For flat damage, $D(S)=1$ for every non-empty $S$, and uniform placement over
$N$ positions, \eqref{eq:payoff} gives $1-c/N$. Fitzsimons and Kashefi prove
$1-\tfrac1{2N}$ for general output and $1-\tfrac1N$ for classical output for
a single trap hidden in an enlarged
graph~\cite[Theorem~8]{fitzsimons_unconditionally_2017}. Their construction
is a single-graph one (Remark~\ref{rem:singlegraph}), not the test-round template used here, so their theorem
is not a special case of Lemma~\ref{lem:reduction}. The coincidence of the
two expressions for $c=\tfrac12$ and $c=1$ is the reason these two values are
used in the examples. For a given trap design the constant $c$ has to be
taken from its soundness proof.
\end{remark}

\begin{remark}[Single-graph constructions]
\label{rem:singlegraph}
In constructions that embed traps and computation in one graph, a qubit that
hosts a trap does not carry computation, so the damage caused by deviating
on it depends on where the trap is. The model above does not apply to such
constructions without modification: one would have to index damage by the
computation slots and detection by the trap slots, which are disjoint in
every instance. We do not pursue this here.
\end{remark}

\subsection{Status of the hypotheses}
\label{sec:model-status}

Hypothesis~\ref{ass:twirl} is a consequence of the client's randomization,
not an assumption about the server, and it is what makes trap analyses
tractable (Section~\ref{sec:prelim-twirl}).
Hypothesis~\ref{ass:blind} combines the blindness of UBQC, which holds for
every distribution of the client's secrets~\cite{broadbent_universal_2009},
with the indistinguishability of test and computation rounds used in
Ref.~\cite{leichtle_verifying_2021}, and with the design choice of revealing
no flag before the end. Hypothesis~\ref{ass:detect} abstracts the per-trap
detection probability. Hypothesis~\ref{ass:damage} is supplied, for patterns
with flow, by the flow cones of Section~\ref{sec:flow}, which are proved
sound and are exact in the instances we simulated.

Hypothesis~\ref{ass:rounds} is a real restriction on the server. It
describes a persistent fault, or a server that sticks to one tampering
pattern, and not a general adversary. A malicious server can deviate in only a few rounds, or
deviate in each round independently with a small probability, and
Theorem~\ref{thm:general} shows that either choice changes the scaling of the
soundness error from exponential to inverse-linear in $T$. Every multi-round
result is therefore stated separately for persistent deviations and for
general servers. If
Hypothesis~\ref{ass:detect} holds only with a smaller constant $c'$, every
guarantee holds with $c'$ in place of $c$. If Hypothesis~\ref{ass:damage}
holds only with a larger function $D'\ge D$, the guarantees hold with $D'$ in
place of $D$ (Proposition~\ref{prop:misspec}).

\subsection{Damage functions}
\label{sec:model-damage}

\begin{definition}[Damage profile and consistent damage functions]
\label{def:damage}
The \emph{damage profile} is $w\in[0,1]^N$ with $w_i=D(\{i\})$. The class
$\cD(w)$ consists of all set functions $D:2^{[N]}\to[0,1]$ with
$D(\{i\})=w_i$ that are \emph{monotone} ($S\subseteq S'$ implies
$D(S)\le D(S')$) and \emph{subadditive} ($D(S\cup S')\le D(S)+D(S')$).
\end{definition}

Monotonicity states that corrupting more positions cannot lower the damage
bound. Subadditivity is the union bound; it holds for every damage notion of
the form ``at least one affected output is corrupted'', including the
flow-cone functions of Section~\ref{sec:flow}. Two members of $\cD(w)$ are
extremal,
\begin{equation}
  \Dmax(S)=\max_{i\in S}w_i,
  \qquad
  \Dadd(S)=\min\Bigl\{1,\;\textstyle\sum_{i\in S}w_i\Bigr\},
  \label{eq:extremal}
\end{equation}
and they bracket the class.

\begin{lemma}[Envelope]
\label{lem:envelope}
For every $D\in\cD(w)$ and every $S$, $\Dmax(S)\le D(S)\le\Dadd(S)$.
\end{lemma}

\begin{proof}
Monotonicity gives $D(S)\ge w_i$ for every $i\in S$. Subadditivity applied
to the singletons of $S$, together with $D\le1$, gives
$D(S)\le\min\{1,\sum_{i\in S}w_i\}$.
\end{proof}

By Lemma~\ref{lem:envelope}, the worst-position model $\Dmax$ is the most
optimistic damage model consistent with $w$, and the capped additive model
$\Dadd$ is the most pessimistic. Results under $\Dmax$ are limits on what
any placement can achieve, and results under $\Dadd$ hold whatever the true
combination rule is.

\section{The Trap Placement Game and its Optimistic Envelope}
\label{sec:game}

\begin{definition}[Trap placement game]
\label{def:game}
For $D\in\cD(w)$ and $c\in(0,1]$, the game $\Gamma_D$ is played as follows.
The client chooses $p\in\Delta_N$. The server chooses a distribution over
non-empty supports $S\subseteq[N]$. Neither observes the other's choice.
The server receives $U_D(p,S)$ from \eqref{eq:payoff} and the client pays
it. The value is $v_D=\min_{p}\max_{S\neq\emptyset}U_D(p,S)$. We write
$\vmax:=v_{\Dmax}$ and $\vrob:=v_{\Dadd}$.
\end{definition}

The payoff is affine in $p$ for each $S$, so $\max_SU_D(p,S)$ is convex in
$p$ and a minimizer exists. Because the payoff is bilinear in the mixed
strategies, the minimax theorem \eqref{eq:minimax}~\cite{vonneumann_theory_1944,sion_general_1958}
guarantees a saddle point.

The game $\Gamma_D$ of Definition~\ref{def:game} is the single-round problem. By
Lemma~\ref{lem:reduction} its value bounds the soundness error of one test
round and one computation round, and it is the $T=1$ case of the persistent
payoff \eqref{eq:payoffT}. Sections~\ref{sec:game}--\ref{sec:robust} study
this problem because it isolates the effect of the combination rule.
Section~\ref{sec:budget} shows how the answer changes with $T$ and for
general servers, where the damage-weighted placement of
Section~\ref{sec:robust} is exactly optimal for every $D\in\cD(w)$.
Proposition~\ref{prop:sandwich} shows that $\vmax$ and $\vrob$ bracket the
value of every game in the family, and Remark~\ref{rem:commit} explains why
the client loses nothing by publishing its placement distribution.

\begin{proposition}[Sandwich]
\label{prop:sandwich}
For every $D\in\cD(w)$, $\vmax\le v_D\le\vrob$. Moreover,
\[
  \min_{p\in\Delta_N}\;\sup_{D\in\cD(w)}\;\max_{S\neq\emptyset}U_D(p,S)=\vrob,
\]
so $\vrob$ is the value of the placement problem that is robust to the
unknown damage model.
\end{proposition}

\begin{proof}
Lemma~\ref{lem:envelope} gives $U_{\Dmax}\le U_D\le U_{\Dadd}$ pointwise,
and minimax values are monotone in the payoff. The supremum over
$D\in\cD(w)$ is attained at $\Dadd\in\cD(w)$ for every $p$ and $S$.
\end{proof}

\begin{remark}[Publishing the placement costs nothing]
\label{rem:commit}
The inner maximization already lets the server best-respond to $p$. The
value $v_D$ is therefore the value of the Stackelberg game in which the
client announces $p$. By the minimax theorem it also equals the
simultaneous-move value, as is standard for zero-sum security
games~\cite{korzhyk_stackelberg_2011}. The client may publish its placement
distribution in keeping with Shannon's maxim that ``the enemy knows the
system''~\cite{shannon_communication_1949}. Only the sampled position must
stay secret, and blindness keeps it secret. We likewise assume that the
server knows $w$, which is conservative because $w$ is computed from the
public graph (Section~\ref{sec:flow}).
\end{remark}

\subsection{Closed-form solution for the optimistic model}

Under $\Dmax$, for every $S$ with $i^\ast\in\arg\max_{i\in S}w_i$ we have
$U_{\Dmax}(p,S)=w_{i^\ast}(1-c\,p(S))\le w_{i^\ast}(1-c\,p_{i^\ast})$. The
server therefore loses nothing by attacking a single position, and
\begin{equation}
  \vmax=\min_{p\in\Delta_N}\max_{i\in[N]}w_i\,(1-c\,p_i).
  \label{eq:program}
\end{equation}
For \emph{any} $D\in\cD(w)$ the server can also restrict itself to
singletons, which is a second way to see that $v_D\ge\vmax$.
Problem~\eqref{eq:program} is a linear program in $(p,v)$ with $N$
constraints, and its solution has the water-filling form familiar from
power allocation~\cite{cover_elements_2006}. The next theorem is
elementary. We state it with a proof because the equilibrium strategy of
the server, the uniqueness of the placement and the structure of the active
set are used in Sections~\ref{sec:robust} and~\ref{sec:budget}.

\begin{theorem}[Water-filling]
\label{thm:minimax}
Order positions so that $w_1\ge\dots\ge w_{N_+}>0=w_{N_++1}=\dots$. For
$k\le N_+$ let $H_k=\sum_{i\le k}w_i^{-1}$ and $v_k=(k-c)/H_k$. Then:
\begin{enumerate}
\item The set $K=\{k\le N_+:\,v_k<w_k\}$ is an initial segment
      $\{1,\dots,k^\ast\}$. Moreover $w_{k^\ast+1}\le v_{k^\ast}$, with the
      convention $w_{N_++1}=0$.
\item The value is $\vmax=v_{k^\ast}$. The unique optimal placement is
      \begin{equation}
        p^\star_i=\frac1c\Bigl(1-\frac{\vmax}{w_i}\Bigr)
        \quad(i\le k^\ast),\qquad p^\star_i=0\quad(i>k^\ast).
        \label{eq:solution}
      \end{equation}
\item The server strategy $\sigma^\star_i=w_i^{-1}/H_{k^\ast}$ on
      $A=\{1,\dots,k^\ast\}$ guarantees $\vmax$, so
      $(p^\star,\sigma^\star)$ is a Nash equilibrium. If
      $w_{k^\ast+1}<\vmax$, then $\sigma^\star$ is the server's unique
      equilibrium strategy.
\end{enumerate}
\end{theorem}

\begin{proof}
\emph{(1)} For $k<N_+$ we have the equivalence
\begin{equation}
  v_{k+1}<w_{k+1}
  \iff k-c<w_{k+1}H_k
  \iff v_k<w_{k+1}.
  \label{eq:equiv}
\end{equation}
Since $c>0$, $v_1=(1-c)w_1<w_1$, so $1\in K$. Suppose $k\notin K$, that is,
$v_k\ge w_k$. Then $v_k\ge w_k\ge w_{k+1}$, and \eqref{eq:equiv} gives
$v_{k+1}\ge w_{k+1}$, so $k+1\notin K$. Hence $K$ is an initial segment.
If $k^\ast<N_+$, then $k^\ast+1\notin K$, and \eqref{eq:equiv} gives
$v_{k^\ast}\ge w_{k^\ast+1}$. If $k^\ast=N_+$ the claim is trivial.

\emph{(2) Feasibility and upper bound.} For $i\le k^\ast$ we have
$\vmax<w_{k^\ast}\le w_i$, so $p^\star_i>0$. Also
$\sum_{i\le k^\ast}p^\star_i=(k^\ast-\vmax H_{k^\ast})/c=1$. For $i\le k^\ast$,
$w_i(1-c\,p^\star_i)=\vmax$. For $i>k^\ast$, $w_i\le w_{k^\ast+1}\le\vmax$ by
part~(1). Hence $\max_iw_i(1-cp^\star_i)=\vmax$.

\emph{Lower bound and uniqueness.} Suppose a placement $p$ achieves
$\max_iw_i(1-cp_i)\le\vmax$. Then $p_i\ge p^\star_i$ for all $i\le k^\ast$.
Since these $p^\star_i$ already sum to one, $p=p^\star$. So no placement does
better, and $p^\star$ is the unique optimum.

\emph{(3)} For any $p$,
$\sum_{i\in A}\sigma^\star_iw_i(1-cp_i)=(k^\ast-c\sum_{i\in A}p_i)/H_{k^\ast}\ge
(k^\ast-c)/H_{k^\ast}=\vmax$. For uniqueness, when $w_{k^\ast+1}<\vmax$
every optimal server strategy $\sigma$ is supported on $A$, because only
positions in $A$ pay $\vmax$ against $p^\star$. Its payoff against the pure
client strategy $e_j$ is $\sum_i\sigma_iw_i-c\sigma_jw_j$, and this must be
at least $\vmax$ for each $j\in A$. Averaging these payoffs with weights
$p^\star_j>0$ gives exactly the payoff against $p^\star$, which equals
$\vmax$. So every term equals $\vmax$, which forces $\sigma_jw_j$ to be
constant on $A$, that is, $\sigma=\sigma^\star$.
\end{proof}

Remark~\ref{rem:equilibrium} shows that the specific mixture in part~(3) of
Theorem~\ref{thm:minimax} matters, and Corollaries~\ref{cor:fk}
and~\ref{cor:limit} record two consequences of the theorem.

\begin{remark}[Server equilibrium]
\label{rem:equilibrium}
Equilibrium requires the server to mix with $\sigma_i\propto w_i^{-1}$, so
that the client is indifferent among active positions. An arbitrary
distribution on $A$ is \emph{not} an equilibrium strategy. For
$w=(1,0.8,0.6,0.4)$ and $c=\tfrac12$ we find $\vmax=2/3$, whereas the
server that is uniform on $A$ guarantees only $0.65$.
\end{remark}

\begin{corollary}[Flat damage recovers the uniform bound]
\label{cor:fk}
If $w_i=1$ for all $i$, then $k^\ast=N$, $p^\star_i=1/N$ and
$\vmax=1-c/N$. By Remark~\ref{rem:calibration} this is the FK single-trap
bound for $c\in\{\tfrac12,1\}$.
\end{corollary}

\begin{proof}
$v_k=(k-c)/k<1$ for every $k$, so $k^\ast=N$. Then $\vmax=1-c/N$ and
$p^\star_i=(1-\vmax)/c=1/N$.
\end{proof}

\begin{corollary}[A universal limit]
\label{cor:limit}
For every placement $p$ and every $D\in\cD(w)$, the soundness bound
$\max_SU_D(p,S)$ is at least $\vmax$. No placement achieves a smaller
soundness bound under any damage model in $\cD(w)$, so $\vmax$ is a limit
on what placement optimization can deliver.
\end{corollary}

\subsection{Failure of the optimistic placement}

By Corollary~\ref{cor:limit}, no placement beats $\vmax$ under any damage
model in $\cD(w)$. The converse fails: the placement that attains $\vmax$
under $\Dmax$ can be very poor under other damage models.
In Equation~\eqref{eq:solution}, positions with $w_i\le\vmax$ receive no
traps. Under $\Dmax$ that allocation is justified, because a deviation on
such a position pays at most $\vmax$. Under a damage model in which harm
accumulates across positions it is not, as the next proposition shows.

\begin{proposition}[Failure of the optimistic placement]
\label{prop:optimism}
Let $A$ be the active set of Theorem~\ref{thm:minimax}. Then
$\max_SU_{\Dadd}(p^\star,S)\ge\min\{1,\sum_{i\notin A}w_i\}$. In particular,
if the abandoned positions carry total weight at least one, the
water-filling placement has robust soundness error $1$.
\end{proposition}

\begin{proof}
If $A=[N]$ the bound is trivial. Otherwise take $S=[N]\setminus A$. Then $p^\star(S)=0$ and $U_{\Dadd}(p^\star,S)=\Dadd(S)$.
\end{proof}

For $w=(1,0.3,\dots,0.3)$ with $N=8$ the water-filling placement puts the
trap on position~$1$ with certainty. A server that deviates on positions
$2$ to $8$ simultaneously is never detected, and under $\Dadd$ it corrupts
the output with probability one. Across $400$ random profiles with
$2\le N\le10$ and $c\in\{\tfrac12,1\}$, the water-filling placement had a
strictly worse robust guarantee than the damage-weighted placement of
Section~\ref{sec:robust} in $333$ cases. This is the evidence for
hypothesis~W3 of Section~\ref{sec:hypotheses}: a guarantee derived under a
single combination rule can be void under another.

\section{The Robust Envelope: Damage-Weighted Placement}
\label{sec:robust}

This section bounds $\vrob$, the value of the placement problem that is
robust to the unknown damage model, and gives three ways to work with it: a
closed-form placement that is optimal to within a factor of two
(Section~\ref{sec:robust-pw}), an exact dual characterization
(Section~\ref{sec:robust-dual}), and an algorithm that computes the optimum
for hundreds of positions (Section~\ref{sec:robust-alg}).

\subsection{The damage-weighted placement}
\label{sec:robust-pw}

Under the capped additive model the server's best response to a fixed
placement includes minimizing $p(S)$ subject to $w(S)\ge1$, which is a
minimum-knapsack problem and NP-hard in
general~\cite{garey_computers_1979}. The exact value $\vrob$ is the
solution of a linear program with $2^N-1$ constraints. The closed-form
placement of Definition~\ref{def:pw} avoids both difficulties;
Lemma~\ref{lem:packing} and Theorem~\ref{thm:robust} give its guarantees.

\begin{definition}[Damage-weighted placement; packing number]
\label{def:pw}
The \emph{damage-weighted placement} is $\pw=w/\|w\|_1$. A \emph{cover} is a
set $S$ with $w(S)\ge1$. The \emph{fractional packing number} $\kappa(w)$ is
the maximum of $\sum_S\lambda_S$ over $\lambda\ge0$ supported on covers,
subject to $\sum_{S\ni i}\lambda_S\le1$ for every position $i$.
\end{definition}

\begin{lemma}[Packing]
\label{lem:packing}
If $\|w\|_1\ge1$, then $\kappa(w)\ge\|w\|_1/(1+\wmax)\ge\|w\|_1/2$.
\end{lemma}

\begin{proof}
Let $W=\|w\|_1$. Lay the positions end to end as half-open arcs $J_i$ of
length $w_i$ on a circle of circumference $W$. For $u\in[0,W)$ let $S_u$ be
the set of positions whose arcs meet the arc $[u,u+1)$ taken modulo $W$.
Since $W\ge1$, that arc does not overlap itself and is contained in
$\bigcup_{i\in S_u}J_i$, so $w(S_u)\ge1$ and $S_u$ is a cover. The map
$u\mapsto S_u$ is piecewise constant with finitely many pieces. Assign
weight $\lambda=\ell/(1+\wmax)$ to $S_u$ for each piece of length $\ell$.
Position $i$ belongs to $S_u$ for a set of $u$ of measure at most $1+w_i$,
so its load is at most $(1+w_i)/(1+\wmax)\le1$, and the total weight is
$W/(1+\wmax)$.
\end{proof}

\begin{theorem}[Robust placement]
\label{thm:robust}
Let $w\in[0,1]^N$ with $W=\|w\|_1\ge\max\{1,2c\}$.
\begin{enumerate}
\item \emph{(Guarantee.)} For every $D\in\cD(w)$,
      $\max_SU_D(\pw,S)\le1-c/W$, with equality whenever $\wmax=1$.
\item \emph{(Converse.)} $\vrob\ge1-c/\kappa(w)\ge1-c(1+\wmax)/W$.
\item \emph{(Near-optimality.)} The robust detection margin satisfies
      $1-\vrob\le(1+\wmax)c/W\le2c/W$, so no placement improves on the
      margin certified by $\pw$ by more than a factor $1+\wmax$.
\item \emph{(Uniform placement.)} If $\wmax=1$, uniform placement has robust
      value exactly $1-c/N\ge1-c/W$.
\end{enumerate}
\end{theorem}

\begin{proof}
\emph{(1)} Fix $S$ and let $x=w(S)$, so that $\pw(S)=x/W$. By
Lemma~\ref{lem:envelope}, $U_D(\pw,S)\le g(x):=\min\{1,x\}(1-cx/W)$. For
$x\ge1$, $g(x)\le1-c/W$. For $x\in(0,1)$, $g(x)=x-cx^2/W$ is increasing,
because $x<1\le W/(2c)$, so $g(x)<g(1)=1-c/W$. If some $w_i=1$ the singleton
$\{i\}$ attains $1-c/W$ already under $\Dmax$.

\emph{(2)} Let $\lambda$ be an optimal fractional packing of covers and let
the server play the cover $S$ with probability $\lambda_S/\kappa$. For every
placement $p$, since $\Dadd(S)=1$ on covers, the expected payoff is
$1-(c/\kappa)\sum_ip_i\sum_{S\ni i}\lambda_S\ge1-c/\kappa$, which lower
bounds $\vrob$. Lemma~\ref{lem:packing} gives the second inequality.

\emph{(3)} Combine (1) and (2).

\emph{(4)} Under uniform placement $U_{\Dadd}(p,S)\le1-c|S|/N\le1-c/N$, with
equality at a singleton of weight one, and $W\le N$.
\end{proof}

Theorem~\ref{thm:robust} is the robust counterpart of the flat bound
$1-c/N$: placing traps in proportion to damage replaces the number of
positions $N$ by the effective size $\|w\|_1$, and the guarantee holds
whatever the combination rule. The factor $1+\wmax$ approaches one when the
damage is spread thinly, which is the case for the fractional profiles of
Section~\ref{sec:flow}, where $\wmax$ can be below $0.1$
(Section~\ref{sec:robust-alg}). In $400$ random
instances the ratio between the optimal robust margin and the margin of
$\pw$ averaged $1.12$ and never exceeded $1.49$.

\begin{table*}[t]
\centering
\caption{Single trap, $N=8$, $c=\tfrac12$. Here $k^\ast$ is the size of the
water-filling active set and $\vmax$ is the optimistic value
(Theorem~\ref{thm:minimax}); ``WF robust'' is the robust soundness error of
the water-filling placement; $\vrob$ is the exact robust optimum from the
linear program over all $255$ supports; $1-c/\|w\|_1$ is the guarantee of
$\pw$ (Theorem~\ref{thm:robust}); and $\kappa$ is the fractional packing
number. Uniform placement has robust value $1-c/N=0.9375$ in every row. The
exponential profile is $w_i=e^{-(i-1)/3}$.}
\label{tab:advantage}
\small
\begin{tabular}{@{}lcccccccc@{}}
\toprule
Profile $w$ & $k^\ast$ & $\vmax$ & WF robust & $\vrob$ & $1-c/\|w\|_1$ & $1-c/\kappa$ & $\|w\|_1$ & $\kappa$\\
\midrule
Uniform, $w_i=1$            & 8 & 0.9375 & 0.9375 & 0.9375 & 0.9375 & 0.9375 & 8.000 & 8.000\\
Two critical, rest $0.5$    & 2 & 0.7500 & 1.0000 & 0.9000 & 0.9000 & 0.9000 & 5.000 & 5.000\\
One critical, rest $0.3$    & 1 & 0.5000 & 1.0000 & 0.8182 & 0.8387 & 0.8182 & 3.100 & 2.750\\
One critical, rest $0.05$   & 1 & 0.5000 & 0.5000 & 0.5000 & 0.6296 & 0.5000 & 1.350 & 1.000\\
Linear, $1.0\to0.3$         & 3 & 0.7438 & 1.0000 & 0.8889 & 0.9038 & 0.8889 & 5.200 & 4.500\\
Exponential, $\tau=3$       & 2 & 0.6261 & 1.0000 & 0.8435 & 0.8477 & 0.8421 & 3.283 & 3.167\\
\bottomrule
\end{tabular}
\end{table*}

\begin{figure*}[t]
\centering
\includegraphics[width=\textwidth]{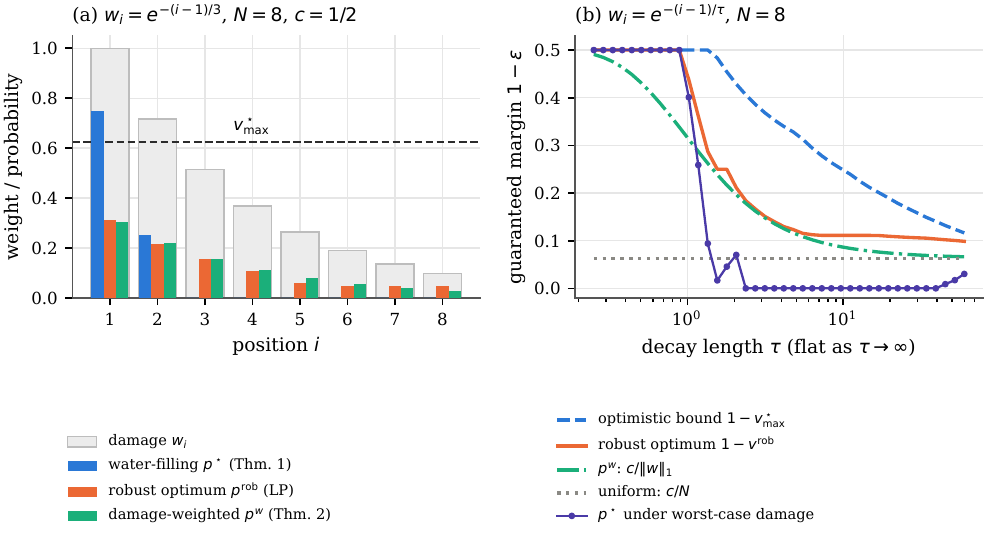}
\caption{(a) For $w_i=e^{-(i-1)/3}$ the water-filling placement (blue) puts
all trap mass on the positions above the threshold $\vmax$ (dashed line).
The exact robust optimum (orange) and the damage-weighted placement (green)
spread mass over all positions and nearly coincide. (b) Certified detection
margin $1-\varepsilon$ as a function of the decay length $\tau$. The
optimistic bound (dashed) cannot be certified without knowing the
combination rule. Evaluated under the worst subadditive damage, the
water-filling placement (violet) loses all margin once the profile is
moderately flat. The damage-weighted placement stays within a factor
$1+\wmax$ of the robust optimum and above uniform placement throughout.}
\label{fig:equilibria}
\end{figure*}

Table~\ref{tab:advantage} shows three regimes. When the positions left
untrapped by the water-filling placement carry total weight at least one,
that placement gives no guarantee against a server that accumulates damage:
its robust soundness error is $1$, even though $\vmax$ is small.
The damage-weighted placement is exactly optimal for the flat and two-level
profiles, and in the remaining rows it is within $0.021$ of the robust
optimum, except for the row ``rest $0.05$'', where its detection margin is a
factor $1.35$ below optimal, still inside the guaranteed factor
$1+\wmax=2$. The packing bound $1-c/\kappa$ is tight in five of the six
rows. When the low-damage positions carry total weight below one, as in the
row ``rest $0.05$'', the abandoned set cannot cause full damage and the
optimistic and robust values coincide. Figure~\ref{fig:equilibria} shows the
three placements and the certified margins as the profile is varied from
concentrated to flat.

\subsection{Damage-load duality}
\label{sec:robust-dual}

The packing bound of Theorem~\ref{thm:robust}(2) uses only covers. The
exact value has a dual description that uses all supports and explains when
the packing bound is tight. Write $\bar D_S:=\Dadd(S)$ for brevity, and for a
server mixture $\sigma$ over non-empty supports define the \emph{expected
damage} and the \emph{damage load} of position $i$,
\begin{equation}
  M(\sigma)=\sum_S\sigma_S\bar D_S,
  \qquad
  L_i(\sigma)=\sum_{S\ni i}\sigma_S\bar D_S .
  \label{eq:loads}
\end{equation}

\begin{theorem}[Damage-load duality]
\label{thm:duality}
For every profile $w$ and every $c\in(0,1]$,
\begin{equation}
  \vrob=\max_{\sigma}\Bigl[M(\sigma)-c\max_{i\in[N]}L_i(\sigma)\Bigr],
  \label{eq:dual}
\end{equation}
the maximum being over probability distributions on non-empty supports.
Equivalently, with $\lambda$ ranging over non-negative measures on
non-empty supports whose damage loads satisfy
$\ell_i(\lambda)=\sum_{S\ni i}\lambda_S\bar D_S\le1$ for all $i$,
\begin{equation}
  \vrob=\max_{\lambda\neq0}\;
  \frac{\sum_S\lambda_S\bar D_S-c}{\sum_S\lambda_S} .
  \label{eq:dual-ratio}
\end{equation}
\end{theorem}

\begin{proof}
The payoff $\sum_S\sigma_S\bar D_S(1-c\,p(S))$ is bilinear in $(p,\sigma)$ on
compact convex sets, so the minimax theorem~\cite{vonneumann_theory_1944}
applies and $\vrob=\max_\sigma\min_p\sum_S\sigma_S\bar D_S(1-c\,p(S))$. For fixed
$\sigma$, $\sum_S\sigma_S\bar D_Sp(S)=\sum_ip_iL_i(\sigma)$, which the client
maximizes by placing all mass on a position of maximal load. This gives
\eqref{eq:dual}. For \eqref{eq:dual-ratio}, given $\sigma$ put
$\lambda=\sigma/L$ with $L=\max_iL_i(\sigma)>0$; then $\ell_i(\lambda)\le1$
and the ratio equals $M(\sigma)-cL$. Conversely, given $\lambda$ put
$\sigma=\lambda/\sum_S\lambda_S$, so that $\max_iL_i(\sigma)\le1/\sum_S\lambda_S$
and the objective of \eqref{eq:dual} is at least the ratio in
\eqref{eq:dual-ratio}.
\end{proof}

Equation~\eqref{eq:dual} has a simple interpretation. The server wants a
large expected damage, spread over many positions so that no single position
carries a large share of it and no single trap is worth much. The client
wants to put its traps where the damage load is largest.
Corollary~\ref{cor:dual-special} recovers the earlier bounds from special
choices of $\lambda$, and Corollary~\ref{cor:certificate} turns
Theorem~\ref{thm:duality} into an optimality certificate.

\begin{corollary}[Bounds from special measures]
\label{cor:dual-special}
(i) Restricting $\lambda$ to covers gives $\vrob\ge1-c/\kappa(w)$.
(ii) Taking $\lambda$ supported on a single set $S$ with $\lambda_S=1/\bar D_S$
gives $\vrob\ge \bar D_S(1-c)$; in particular $\vrob\ge\wmax(1-c)$.
\end{corollary}

\begin{proof}
(i) On covers $\bar D_S=1$, the load constraint becomes the packing constraint,
and the ratio in \eqref{eq:dual-ratio} equals $(\kappa-c)/\kappa$.
(ii) The loads are $\ell_i=1$ for $i\in S$, and the ratio is
$(1-c)/(1/\bar D_S)$. Take $S=\{i\}$ with $w_i=\wmax$.
\end{proof}

\begin{corollary}[Optimality certificate]
\label{cor:certificate}
If a placement $p$ and a measure $\lambda$ with loads at most one satisfy
$\max_SU_{\Dadd}(p,S)=(\sum_S\lambda_S\bar D_S-c)/\sum_S\lambda_S$, then $p$ is
an optimal robust placement and $\lambda$ is an optimal dual measure.
\end{corollary}

We use Corollary~\ref{cor:certificate} to solve a family of profiles in
closed form.

\subsection{Two-level profiles}
\label{sec:robust-twolevel}

A profile is \emph{two-level} when $a$ positions have weight $1$ and the
remaining $b$ positions have a common weight $\delta\in(0,1]$. Such profiles
arise when a computation has a set of critical positions and a background of
positions of equal, smaller influence. Let $k=\lceil1/\delta\rceil$ be the
number of small positions needed to reach full damage.

\begin{proposition}[Two-level profiles]
\label{prop:twolevel}
Let $k\le b$, $\kappa=a+b/k$ and let $p^{\mathrm{cb}}$ be the
\emph{cover-balanced} placement that assigns $1/\kappa$ to each unit-weight
position and $1/(k\kappa)$ to each small position. If
\begin{equation}
  \max_{1\le j<k}\;j\delta\Bigl(1-\frac{cj}{k\kappa}\Bigr)\;\le\;1-\frac c\kappa,
  \label{eq:twolevel-cond}
\end{equation}
then $p^{\mathrm{cb}}$ is an optimal robust placement and
$\vrob=1-c/\kappa$.
\end{proposition}

\begin{proof}
Upper bound. Let $S$ contain $i$ unit positions and $j$ small ones. If
$i\ge1$ then $p^{\mathrm{cb}}(S)\ge1/\kappa$ and $\bar D_S=1$, so the payoff is at
most $1-c/\kappa$. If $i=0$ and $j\ge k$ then $\bar D_S=1$ and
$p^{\mathrm{cb}}(S)\ge k/(k\kappa)=1/\kappa$, with the same conclusion. If
$i=0$ and $j<k$ then the payoff is $j\delta(1-cj/(k\kappa))$, which is at
most $1-c/\kappa$ by \eqref{eq:twolevel-cond}.

Lower bound. Put $\lambda=1$ on each singleton of unit weight, and
distribute weight $(b/k)\binom{b}{k}^{-1}$ on every $k$-subset of the small
positions. All these sets are covers, so $\bar D_S=1$. The load of a unit
position is $1$, and the load of a small position is
$(b/k)\binom{b}{k}^{-1}\binom{b-1}{k-1}=1$. The total mass is
$\sum_S\lambda_S=a+b/k=\kappa$, so \eqref{eq:dual-ratio} gives
$\vrob\ge(\kappa-c)/\kappa$. The two bounds meet, and
Corollary~\ref{cor:certificate} applies.
\end{proof}

Condition~\eqref{eq:twolevel-cond} fails when a set of $k-1$ small positions
carries nearly full damage while attracting little trap mass. In that regime
the optimum is strictly above $1-c/\kappa$ and is obtained from the
algorithm of the next subsection. Over the $430$ two-level instances with
$a\le3$, $b\le9$, $c\in\{0.25,0.5\}$ and twelve values of $\delta$ that
satisfy \eqref{eq:twolevel-cond}, the closed form agreed with the exact
linear program in every case. Proposition~\ref{prop:twolevel} accounts for the
second and third rows of Table~\ref{tab:advantage}, where $\kappa=5$ and
$\kappa=2.75$ and condition \eqref{eq:twolevel-cond} holds. The fourth row
lies outside its hypothesis, since $k=20$ exceeds $b=7$ and the small
positions cannot form a cover; there the optimistic and robust values
coincide.

\subsection{Computing the robust optimum}
\label{sec:robust-alg}

For profiles without closed-form structure, the robust optimum can still be
computed without listing all supports. Round the profile up to a
grid, $w^K_i=\lceil Kw_i\rceil/K$. Since $w\le w^K$ pointwise,
every guarantee proved for $w^K$ is valid for $w$ by
Proposition~\ref{prop:misspec}(2) with $\eta=0$. Over the grid, the
server's best response is a knapsack problem that dynamic programming solves
exactly.

\begin{algorithm}[t]
\caption{Robust placement by column generation}
\label{alg:colgen}
\begin{algorithmic}[1]
\Require profile $w$, constant $c$, grid size $K$
\State $w^K_i\gets\lceil Kw_i\rceil/K$ for all $i$; $\;\mathcal S\gets\{\{\arg\max_iw_i\}\}$
\Loop
  \State solve $\min\{v:\;\bar D_S(1-c\,p(S))\le v\ \forall S\in\mathcal S,\ p\in\Delta_N\}$
  \State let $(p,v)$ be the solution
  \State $B\gets$ best response to $p$ under $w^K$ \Comment{knapsack DP}
  \If{$U(p,B)\le v$} \Return $(p,v)$ \EndIf
  \State $\mathcal S\gets\mathcal S\cup\{B\}$
\EndLoop
\end{algorithmic}
\end{algorithm}

\begin{proposition}[Correctness and cost]
\label{prop:colgen}
Let $w^K$ be the rounded profile. (i) The best response
$\max_SD_{\mathrm{add}}^{w^K}(S)(1-c\,p(S))$ can be computed exactly in
$O(NK)$ time and $O(K)$ space by dynamic programming over the achievable
rounded damage levels. (ii) Algorithm~\ref{alg:colgen} terminates after
finitely many iterations and returns the exact robust optimum of $w^K$.
(iii) The placement it returns certifies soundness error at most
$v^{\mathrm{rob}}(w^K)$ for the true profile $w$, and
$\|w^K\|_1\le\|w\|_1+N/K$.
\end{proposition}

\begin{proof}
(i) Let $W_i=Kw^K_i\in\mathbb N$. For $t=0,\dots,K$ let $\mu[t]$ be the
minimum of $p(S)$ over sets $S$ whose rounded damage $\sum_{i\in S}W_i$,
capped at $K$, equals $t$. The standard knapsack recursion computes $\mu$ in
$O(NK)$ time. The best response value is
$\max_{1\le t\le K}(t/K)(1-c\,\mu[t])$ with $t=K$ interpreted as full
damage, and the maximizing set is recovered from the recursion.
(ii) Each iteration adds a support that is violated by the current
solution, hence not already present, and there are finitely many supports;
on termination the current $p$ satisfies all constraints, so $v$ is the
optimum of the full program.
(iii) Since $w\le w^K$ we have $D\le D_{\mathrm{add}}^{w}\le
D_{\mathrm{add}}^{w^K}$ for every $D\in\cD(w)$, so the payoff under the true
profile is bounded by the payoff under $w^K$. The last claim follows from
$\lceil Kw_i\rceil\le Kw_i+1$.
\end{proof}

By Proposition~\ref{prop:colgen}, Algorithm~\ref{alg:colgen} solves the
single-round problem \eqref{eq:payoff} exactly. With $T>1$ test rounds the robust optimum depends on $T$
(Section~\ref{sec:budget-persistent}), so its output carries no
multi-round optimality guarantee, and the protocol of
Section~\ref{sec:protocol} does not use it.

\begin{figure}[t]
\centering
\includegraphics[width=\columnwidth]{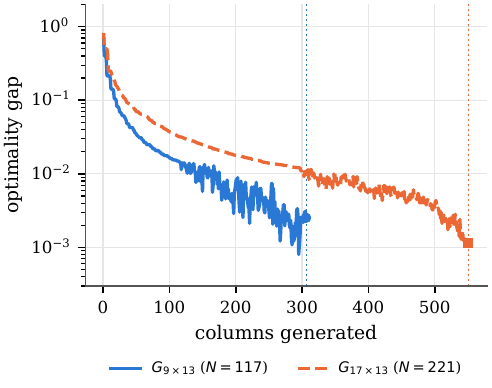}
\caption{Convergence of Algorithm~\ref{alg:colgen} on the fractional
brickwork profiles of Section~\ref{sec:flow}. The optimality gap is the
difference between the best response to the current placement and the value
of the restricted linear program; it is zero at termination, marked by the
dotted verticals. The instances have $N=117$ and $N=221$ positions, for
which the direct formulation would have $2^{117}-1$ and $2^{221}-1$
constraints.}
\label{fig:colgen}
\end{figure}

On random instances with $N\le12$ and $K=2000$, Algorithm~\ref{alg:colgen}
reproduced the brute-force optimum to $3\times10^{-16}$ in all $40$ cases
tested. Table~\ref{tab:colgen} applies it to the fractional brickwork
profiles of Section~\ref{sec:flow}, which have too many supports for direct
enumeration. Figure~\ref{fig:colgen} shows the convergence on the two smaller
instances: the optimality gap reaches zero after $307$ and $551$ generated
columns. These profiles take values in $\{0,\tfrac1n,\dots,1\}$, so with
$K$ a multiple of $n$ the rounding is exact and the reported $\vrob$ is the
robust optimum of the true profile. The robust values of $\pw$ and of
uniform placement in the table are computed exactly with the same knapsack
oracle. The damage-weighted placement is within $2.4\%$ of the optimal
detection margin, and exactly optimal for $G_{33\times13}$, where
$\wmax=0.18$; this is the behavior predicted by the factor $1+\wmax$ of
Theorem~\ref{thm:robust}. Uniform placement loses a factor between $1.4$ and
$1.6$ in these instances. The loss is moderate because a cover of a
fractional profile needs several positions, which blunts the advantage of
concentrating traps; for the binary profiles of Section~\ref{sec:flow} the
loss is the ratio $N/|B|$, which grows linearly in the width.

\begin{table}[t]
\centering
\caption{Robust optimum for brickwork fractional damage profiles computed by
Algorithm~\ref{alg:colgen} with $c=\tfrac12$ and $K=n\lceil400/n\rceil$, for
which the rounding is exact. Margins are $1-\varepsilon$, and
$v(\cdot)$ denotes the exact robust value of a placement. The last two
columns compare the margin of the damage-weighted placement with the
optimum and with uniform placement.}
\label{tab:colgen}
\small
\setlength{\tabcolsep}{3.5pt}
\begin{tabular}{@{}lcccccc@{}}
\toprule
Pattern & $N$ & $\|w\|_1$ & columns & $\vrob$ & $\frac{1-\vrob}{1-v(\pw)}$ & $\frac{1-v(\pw)}{1-v(\text{unif})}$\\
\midrule
$G_{9\times13}$  & 117 & 42.78 & 307  & 0.98804 & 1.023 & 1.37\\
$G_{17\times13}$ & 221 & 45.71 & 551  & 0.98900 & 1.006 & 1.61\\
$G_{33\times13}$ & 429 & 47.30 & 1187 & 0.98943 & 1.000 & 1.51\\
\bottomrule
\end{tabular}
\end{table}

\subsection{Misspecified profiles}
\label{sec:misspec}

The client usually knows $w$ only approximately, and the two envelopes react
to estimation errors in different ways.

\begin{proposition}[Misspecification]
\label{prop:misspec}
\begin{enumerate}
\item \emph{(Optimistic model, additive error.)} Let $\hat w$ be an estimate
      normalized so that $\max_i\hat w_i=1$, and let $\hat p^\star$ be its
      water-filling placement. Then
      $\max_iw_i(1-c\hat p^\star_i)\le\vmax(\hat w)+\|w-\hat w\|_\infty$.
\item \emph{(Robust model, multiplicative error.)} If
      $w_i\le(1+\eta)\hat w_i$ for all $i$ and
      $(1+\eta)\|\hat w\|_1\ge\max\{1,2c\}$, then the placement $p^{\hat w}$
      satisfies $\max_SU_D(p^{\hat w},S)\le1-c/((1+\eta)\|\hat w\|_1)$ for
      every $D\in\cD(w)$. In particular, over-estimating damage, that is
      $\eta=0$ with $\hat w\ge w$, never invalidates the guarantee.
\end{enumerate}
\end{proposition}

\begin{proof}
(1) Write $\delta=\|w-\hat w\|_\infty$. Then
$w_i(1-c\hat p^\star_i)\le\hat w_i(1-c\hat p^\star_i)+\delta$, and the first
term is at most $\vmax(\hat w)$ by Theorem~\ref{thm:minimax}.
(2) Put $W'=(1+\eta)\|\hat w\|_1$. For any $S$ with $x=w(S)$ we have
$p^{\hat w}(S)=\hat w(S)/\|\hat w\|_1\ge x/W'$, and the argument of
Theorem~\ref{thm:robust}(1) applies with $W'$ in place of $W$.
\end{proof}

Part~(1) is stated relative to $\vmax(\hat w)$, a quantity the client can
compute; the analogous bound relative to $\vmax(w)$ is false in general,
because the active set can change discontinuously. Numerically, for
$w=(1,0.85,0.7,0.5,0.35,0.2)$ with each weight perturbed uniformly by at
most $\pm\varepsilon$ before optimizing, the mean realized optimistic value
is $0.693$, $0.729$, $0.765$ and $0.815$ for $\varepsilon=0$, $0.1$, $0.2$
and $0.35$, against $0.917$ for uniform placement; the worst draws reach
$0.802$, $0.933$ and $1.000$, so on unfavorable draws the optimistic
placement can be worse than uniform. Part~(2) has no such failure mode,
which is one reason to prefer $\pw$ when the profile is uncertain. It also
allows the client to use any over-approximation of the damage, such as the
flow cones of Section~\ref{sec:flow} or the rounded profile of
Algorithm~\ref{alg:colgen}.

\section{Repetition: Persistent Deviations and General Servers}
\label{sec:budget}

With $T$ test rounds the soundness error depends on how the server spreads
its deviation over the rounds. Section~\ref{sec:budget-persistent} treats
persistent deviations (Hypothesis~\ref{ass:rounds}), for which the payoff is
$D(S)(1-c\,p(S))^T$ by Lemma~\ref{lem:reduction} and decays exponentially
in $T$. Section~\ref{sec:budget-general} removes that hypothesis. The decay
then becomes inverse-linear, and the placement problem reduces to minimizing
the ratio of damage to trap coverage.

\subsection{Persistent deviations}
\label{sec:budget-persistent}

The optimistic problem remains tractable.

\begin{proposition}[Optimistic envelope, $T$ test rounds]
\label{prop:multitrap}
Let $\vmax(T)=\min_p\max_iw_i(1-cp_i)^T$. The function
$\Phi(v)=\sum_i\bigl(1-(v/w_i)^{1/T}\bigr)_+$ is continuous and strictly
decreasing on $(0,\wmax)$, and $\vmax(T)$ is the unique root of $\Phi(v)=c$
in $[\wmax(1-c)^T,\wmax)$. The unique optimal placement is
$p^\star_i(T)=\frac1c(1-(\vmax/w_i)^{1/T})_+$. For flat damage,
$p^\star=1/N$ and $\vmax(T)=(1-c/N)^T$.
\end{proposition}

\begin{proof}
At $v=\wmax(1-c)^T$ the term for the largest weight equals $c$, so
$\Phi\ge c$. As $v\to\wmax$, $\Phi\to0$. Monotonicity gives a unique root.
The placement $p^\star(T)$ sums to one and equalizes $w_i(1-cp_i)^T=\vmax$
on $\{w_i>\vmax\}$, while every other position pays at most $\vmax$. If some
$p$ achieved a smaller maximum, then $p_i>p^\star_i$ on the active set,
which contradicts normalization. The flat case follows by symmetry.
\end{proof}

As $T$ grows, $\vmax(T)\to0$. The active set then expands to all positions
with $w_i>0$, and $p^\star(T)$ approaches the uniform distribution on them.
Leaving positions untrapped, which is what causes the failure in
Proposition~\ref{prop:optimism}, therefore happens only at small budgets.
The next theorem makes this precise.

\begin{theorem}[Collapse of the envelopes]
\label{thm:collapse}
Suppose $\wmin>0$ and $\vmax(T)\le\wmin/2$. Then for every $D\in\cD(w)$,
$\max_SD(S)(1-c\,p^\star(T)(S))^T=\vmax(T)$. Consequently
$\vrob(T)=\vmax(T)$, and the water-filling placement is robustly optimal.
The condition holds whenever
\begin{equation}
  T\;\ge\;\frac{\ln(2/\wmin)}{\ln\bigl(1/(1-c/N)\bigr)},
  \label{eq:Tcollapse}
\end{equation}
and in particular whenever $T\ge(N/c)\ln(2/\wmin)$.
\end{theorem}

\begin{proof}
Write $v=\vmax(T)$ and $p=p^\star(T)$. Since $v<\wmin$, every position is
active and $(1-cp_i)^T=v/w_i$. For $|S|=s\ge2$, the Weierstrass product
inequality $\prod_i(1-a_i)\ge1-\sum_ia_i$, valid for $a_i\in[0,1]$, gives
$0\le1-c\,p(S)\le\prod_{i\in S}(1-cp_i)$. Hence
\begin{align*}
  \Dadd(S)(1-c\,p(S))^T
  &\le\Bigl(\sum_{j\in S}w_j\Bigr)\prod_{i\in S}\frac{v}{w_i}\\
  &= v\sum_{j\in S}\prod_{i\in S\setminus\{j\}}\frac v{w_i}\\
  &\le v\,s\,\Bigl(\frac{v}{\wmin}\Bigr)^{s-1}\le v\,s\,2^{1-s}\le v .
\end{align*}
Lemma~\ref{lem:envelope} extends this bound to every $D\in\cD(w)$, and
singletons attain $v$. For \eqref{eq:Tcollapse}, uniform placement is
feasible, so $\vmax(T)\le\wmax(1-c/N)^T\le(1-c/N)^T$. Finally,
$\ln(1/(1-x))\ge x$.
\end{proof}

At small $T$, by contrast, the damage-weighted placement keeps its robust
character, but its advantage fades as the budget grows.

\begin{proposition}[$\pw$ with $T$ test rounds]
\label{prop:pwT}
If $W=\|w\|_1\ge1$, then for every $D\in\cD(w)$,
$\max_SD(S)(1-c\,\pw(S))^T\le\max\{(1-c/W)^T,\;W/(c(T+1))\}$.
\end{proposition}

\begin{proof}
With $x=w(S)$, the payoff is at most $\min\{1,x\}(1-cx/W)^T$. For $x\ge1$
this is at most $(1-c/W)^T$. For $x<1$, the function $x(1-cx/W)^T$ is
maximized at $x_0=W/(c(T+1))$, where its value is at most $x_0$. If
$x_0\ge1$, the function is increasing on $(0,1)$, and the first bound
applies.
\end{proof}

\begin{figure}[t]
\centering
\includegraphics[width=\columnwidth]{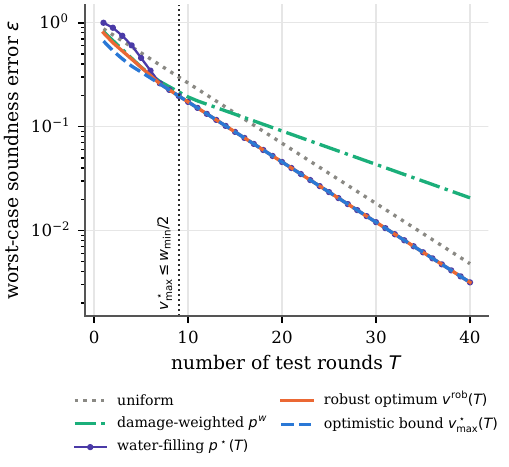}
\caption{Robust soundness error for persistent deviations against the number of test
rounds for the profile of
Table~\ref{tab:budget}. The optimistic bound (dashed) and the robust optimum
(solid orange) separate at small $T$ and merge at $T=8$. The dotted
vertical line marks the sufficient condition of
Theorem~\ref{thm:collapse}. Past it the water-filling placement is robustly
optimal, while $\pw$ decays more slowly.}
\label{fig:budget}
\end{figure}

The second term shows that $\pw$ guarantees only polynomial decay
uniformly over profiles, because a server that deviates on a low-weight set
causes little damage but is rarely caught. Table~\ref{tab:budget} and
Figure~\ref{fig:budget} show both regimes for $w=(1,0.8,0.6,0.4)$. For
$T\le5$ the water-filling placement is far from robust, and the exact robust
optimum, computed by bisection over linear feasibility programs, is close to
the value of $\pw$. From $T=8$ the robust optimum coincides with
$\vmax(T)$; Theorem~\ref{thm:collapse} certifies this from $T=9$, and at
$T=20$ uniform placement already outperforms $\pw$. The resulting rule is to
use $\pw$ when the budget is small or the profile is uncertain, and to
switch to water-filling once \eqref{eq:Tcollapse}, or its sharper form
$\vmax(T)\le\wmin/2$, holds. The single-round optimum of
Algorithm~\ref{alg:colgen} is not part of this rule: Table~\ref{tab:budget}
shows that the optimal placement changes with $T$, so a multi-round
optimization would need its own computation and its own guarantee.

\begin{table*}[t]
\centering
\caption{Worst-case soundness error for persistent deviations against the number of
test rounds $T$, for $w=(1,0.8,0.6,0.4)$ and $c=\tfrac12$. The columns give the
optimistic value $\vmax(T)$; the robust optimum $\vrob(T)$ (exact, all 15
supports); and the robust error of the water-filling placement
$p^\star(T)$, of $\pw$, and of uniform placement. Theorem~\ref{thm:collapse}
guarantees $\vrob=\vmax$ for $T\ge9$.}
\label{tab:budget}
\small
\begin{tabular}{@{}rcccccccc@{}}
\toprule
$T$ & 1 & 2 & 3 & 5 & 8 & 12 & 20 & 40\\
\midrule
$\vmax(T)$                  & 0.6667 & 0.5378 & 0.4498 & 0.3355 & 0.2258 & 0.1327 & 0.0457 & 0.0032\\
$\vrob(T)$                  & 0.8000 & 0.6429 & 0.5362 & 0.3727 & 0.2258 & 0.1327 & 0.0457 & 0.0032\\
water-filling, robust       & 1.0000 & 0.8963 & 0.7497 & 0.4588 & 0.2258 & 0.1327 & 0.0457 & 0.0032\\
$\pw$, robust               & 0.8214 & 0.6747 & 0.5543 & 0.3740 & 0.2423 & 0.1644 & 0.0909 & 0.0206\\
uniform, robust             & 0.8750 & 0.7656 & 0.6699 & 0.5129 & 0.3436 & 0.2014 & 0.0692 & 0.0048\\
\bottomrule
\end{tabular}
\end{table*}

\begin{corollary}[Test rounds needed, persistent deviations]
\label{cor:resources}
Under Hypothesis~\ref{ass:rounds}, the placement $p$ reaches soundness
error $\varepsilon$ with
$T_\varepsilon=\min\{T:\max_SD(S)(1-c\,p(S))^T\le\varepsilon\}$ test
rounds. For a binary profile supported on $B\subseteq[N]$, placing traps
uniformly on $B$ gives
$T_\varepsilon=\lceil\ln(1/\varepsilon)/\ln(1/(1-c/|B|))\rceil$, and
uniform placement over all positions gives the same expression with $N$ in
place of $|B|$. The ratio of the two tends to $N/|B|$ as $c/|B|\to0$.
\end{corollary}

\subsection{General servers}
\label{sec:budget-general}

Without Hypothesis~\ref{ass:rounds}, the server may deviate differently in
different rounds. By Hypothesis~\ref{ass:blind}, its sequence of deviations
$(E_1,\dots,E_{T+1})$ is independent of the index $\kappa$ of the
computation round and of the trap positions. Write $S_r=\supp E_r$, with
$S_r=\emptyset$ when the server is honest in round $r$. For a placement $p$
and a damage function $D$ define the \emph{damage-to-coverage ratio}
\begin{equation}
  \rho_D(p)=\max_{S:\,D(S)>0}\frac{D(S)}{p(S)},
  \label{eq:rho}
\end{equation}
with $\rho_D(p)=\infty$ if some $S$ with $D(S)>0$ has $p(S)=0$.

\begin{theorem}[General servers]
\label{thm:general}
Assume Hypotheses~\ref{ass:twirl}--\ref{ass:damage}.
\begin{enumerate}
\item \emph{(Upper bound.)} For every server strategy,
      \begin{equation}
        \Pr[\text{accept}\wedge\text{incorrect}]\le
        \min\Bigl\{1,\;\frac{\rho_D(p)}{c\,(T+1)}\Bigr\}.
        \label{eq:general-ub}
      \end{equation}
\item \emph{(Optimal placement.)} For every $D\in\cD(w)$,
      $\min_p\rho_D(p)=\|w\|_1$, and the minimum is attained by $\pw$ and by
      no other placement.
\item \emph{(Tightness.)} Suppose the damage and detection bounds of
      Hypotheses~\ref{ass:detect} and~\ref{ass:damage} are attained on
      singletons. For every placement $p$ and every position $i$ with
      $c\,p_i\le\frac12$ and $T+1\ge1/(c\,p_i)$, there is a server strategy
      with
      \begin{equation}
        \Pr[\text{accept}\wedge\text{incorrect}]\ge\frac{w_i}{4c\,p_i\,(T+1)}.
        \label{eq:general-lb}
      \end{equation}
\end{enumerate}
\end{theorem}

\begin{proof}
(1) Condition on the deviation sequence. The computation round is
$\kappa$ with probability $1/(T+1)$; it is incorrect with probability at
most $a_\kappa:=D(S_\kappa)$, and each other round, being a test round with
an independent trap, fails to flag with probability at most
$b_r:=1-c\,p(S_r)$. Hence
\[
  \Pr[\text{accept}\wedge\text{incorrect}\mid(S_r)]\le
  \frac1{T+1}\sum_\kappa a_\kappa\prod_{r\ne\kappa}b_r .
\]
By definition of $\rho=\rho_D(p)$, $a_r\le\rho\,p(S_r)=(\rho/c)(1-b_r)$ for
every $r$, including $S_r=\emptyset$. Therefore
\[
  \sum_\kappa a_\kappa\prod_{r\ne\kappa}b_r\le\frac\rho c
  \sum_\kappa(1-b_\kappa)\prod_{r\ne\kappa}b_r\le\frac\rho c ,
\]
because the middle sum is the probability that exactly one of $T+1$
independent events with probabilities $1-b_r$ occurs. The bound by $1$ is
trivial, and averaging over the sequence gives \eqref{eq:general-ub}.

(2) For every $p$ and every $D\in\cD(w)$, restricting \eqref{eq:rho} to
singletons gives $\rho_D(p)\ge\max_{i:w_i>0}w_i/p_i$. Since
$\sum_{i:w_i>0}p_i(w_i/p_i)=\|w\|_1$ and $\sum_{i:w_i>0}p_i\le1$, this maximum
is at least $\|w\|_1$, with equality only if $w_i/p_i$ is constant on
$\{w_i>0\}$ and $p$ vanishes elsewhere, that is, only if $p=\pw$.
Conversely, for $p=\pw$ and every $S$, Lemma~\ref{lem:envelope} gives
$D(S)/\pw(S)\le\min\{1,w(S)\}\,\|w\|_1/w(S)\le\|w\|_1$.

(3) Let the server deviate on $\{i\}$ in $k=\lceil1/(c\,p_i)\rceil\le T+1$
fixed rounds and be honest in the others. The output is accepted and
incorrect exactly when $\kappa$ is one of the $k$ rounds, the computation is
corrupted, and none of the other $k-1$ deviating rounds, which are test
rounds, flags. Under the stated attainment this has probability
$\frac{k}{T+1}\,w_i\,(1-c\,p_i)^{k-1}\ge\frac{w_i}{c\,p_i(T+1)}\,
(1-c\,p_i)^{1/(c\,p_i)}$, and $(1-x)^{1/x}\ge\frac14$ for $x\le\frac12$.
\end{proof}

Theorem~\ref{thm:general} changes the picture in three ways. First, the
soundness error of a protocol with one computation round decays only as
$1/T$ against a general server, as expected: a server that deviates in few
rounds is rarely tested. Deviating in every round independently with a
small probability has the same effect: with probability
$\alpha=1/(c\,p_i(T+1))$ per round on $\{i\}$ the success probability is
$\alpha w_i(1-\alpha c\,p_i)^T\ge w_i/(e\,c\,p_i(T+1))$, so exponential decay
requires persistence and not only independence across rounds. Second, the
placement problem is solved exactly: the damage-weighted placement minimizes
the bound for every damage function in $\cD(w)$ simultaneously, so the
distinction between the optimistic and the robust envelope disappears.
Third, the gain over uniform placement is the same factor as in the
single-round analysis. For $\pw$ the bound is $\|w\|_1/(c(T+1))$ and, when
some $w_i=1$, $\|w\|_1\ge2c$ and $T+1\ge\|w\|_1/c$, the lower bound
\eqref{eq:general-lb} of part~(3) becomes $\|w\|_1/(4c(T+1))$. For uniform placement the same argument gives
$N/(c(T+1))$ and $N/(4c(T+1))$. Both quantities therefore scale as
$\Theta(\rho/(cT))$, with $\rho=\|w\|_1$ against $\rho=N$.

\begin{corollary}[Test rounds needed, general servers]
\label{cor:resources-general}
Against general servers, $T+1\ge\rho_D(p)/(c\,\varepsilon)$ test rounds
suffice for soundness error $\varepsilon$. For $\pw$ this is
$\|w\|_1/(c\,\varepsilon)$, and for a binary profile supported on $B$ it is
$|B|/(c\,\varepsilon)$, against $N/(c\,\varepsilon)$ for uniform placement.
\end{corollary}

\begin{figure}[t]
\centering
\includegraphics[width=\columnwidth]{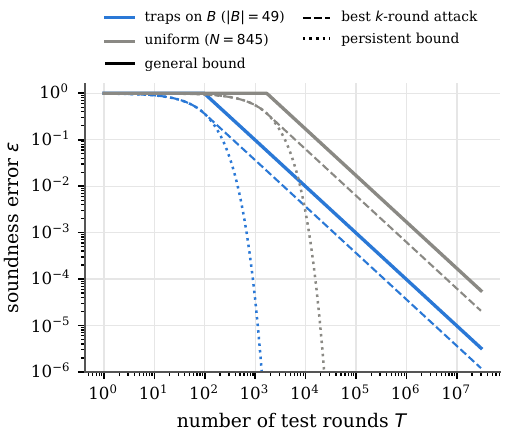}
\caption{Soundness error against the number of test rounds for the decision
qubit of the brickwork pattern $G_{65\times13}$ with $c=\tfrac12$. Blue:
traps uniform on the flow cone $B$ of the decision qubit ($|B|=49$), which is
the damage-weighted placement for this binary profile. Gray: traps uniform
over all $N=845$ positions. Solid lines are the general-server bound of
Theorem~\ref{thm:general}(1), dashed lines the best $k$-round attack on a
single position, and dotted lines the persistent-deviation bound of
Lemma~\ref{lem:reduction}.}
\label{fig:regimes}
\end{figure}

Figure~\ref{fig:regimes} compares the two regimes for the brickwork decision
qubit of Section~\ref{sec:flow}. Against persistent deviations about $674$
test rounds reach $\varepsilon=10^{-3}$ with traps on the flow cone, against
$11\,671$ with uniform placement. Against general servers the corresponding
sufficient numbers are $9.8\times10^4$ and $1.69\times10^6$. The ratio is
about $17$ in both regimes. The absolute numbers for
general servers show why practical protocols use many computation rounds
and a majority vote~\cite{leichtle_verifying_2021}; how non-uniform placement
enters such protocols is discussed in Section~\ref{sec:discussion}.

\section{Damage Profiles from the Flow: Brickwork Patterns}
\label{sec:flow}

The results above take a damage profile as input. This section shows how to
compute one from the pattern using only public information, proves that it
is sound, and checks it against exact simulation.

\subsection{Flow cones}

Consider a pattern on an open graph $(G,I,O)$ with flow
$(f,\prec)$~\cite{danos_determinism_2006,danos_measurement_2007}. By
\eqref{eq:adapt}, the outcome $s_u$ of a measured qubit $u$ enters the
adapted angles, or the final corrections, only of $f(u)$ and of the qubits
in $\mathcal N(f(u))\setminus\{u\}$, all of which come after $u$. The
\emph{dependency DAG} has an arc from $u$ to each of these qubits. The
\emph{flow cone} $F(u)$ is the set of vertices reachable from $u$ in this
DAG, including $u$, and $F(S)=\bigcup_{u\in S}F(u)$.

\begin{proposition}[Soundness of flow cones for brickwork patterns]
\label{prop:influence}
Consider a brickwork pattern with the row flow $f(i,j)=(i,j+1)$, executed
with the client's pads. Let a Pauli operator supported on $S$ act after the
entangling operations and before the measurements, as in
Hypothesis~\ref{ass:twirl}. Then, for every output qubit $o\notin F(S)$, the
reduced state of $o$ after the client's corrections equals its ideal
reduced state. Consequently, for a decision register $O^\star\subseteq O$,
\begin{equation}
  D_{\mathrm{bin}}(S)=\1\bigl[F(S)\cap O^\star\neq\emptyset\bigr],\quad
  D_{\mathrm{frac}}(S)=\frac{|F(S)\cap O|}{|O|}
  \label{eq:flowdamage}
\end{equation}
are monotone subadditive damage functions. $D_{\mathrm{bin}}$ bounds the
probability of the incorrectness event of Section~\ref{sec:model-template}.
$D_{\mathrm{frac}}$ bounds the average over output qubits of the trace
distance between the actual and the ideal single-qubit output states.
\end{proposition}

\begin{proof}
\emph{Step 1: each Pauli acts just before its own measurement.} A Pauli
factor on qubit $v$ commutes with the measurements of all other qubits, and
the client's classical processing acts only on outcomes. Moving every factor
of the deviation to the moment immediately before its own qubit is measured,
or before the output corrections for an output qubit, therefore changes
neither the joint distribution of outcomes nor the final state.

\emph{Step 2: a Pauli before a measurement is a modified measurement.} For
the basis $\ket{\pm_\delta}$ one has $X\ket{\pm_\delta}\propto\ket{\pm_{-\delta}}$
and $Z\ket{\pm_\delta}=\ket{\mp_\delta}$. Hence $X$ before the measurement of
$v$ is the same as measuring $v$ at angle $-\delta_v$, $Z$ is the same as
reporting the complementary outcome, and $Y$ combines both. After the
measurement the qubit is discarded, so nothing else changes. On an output
qubit the Pauli remains on that qubit.

\emph{Step 3: a reported outcome flip toggles only corrections in the DAG.} If
the reported outcome of $v$ is flipped, the client applies the corrections
for the wrong value. By \eqref{eq:adapt} this toggles the sign of the angle
of $f(v)$ and adds $\pi$ to the angles of $\mathcal N(f(v))\setminus\{v\}$,
or toggles an $X$ or $Z$ correction when these are outputs. By Step~2, each
toggle is again a modified measurement at a DAG successor of $v$. By
induction over the measurement order, every deviation supported on $S$ is
equivalent to running the pattern with modified angles on a set of measured
qubits contained in $F(S)$, together with Pauli operators on output qubits in
$F(S)$.

\emph{Step 4: modified angles are local gate changes.} A pattern with flow
is deterministic for every choice of angles~\cite{danos_determinism_2006}.
Changing the angles on a set $A$ therefore yields the circuit of the ideal
pattern with the single-qubit gates of the steps in $A$ replaced. For the
brickwork pattern this circuit is explicit~\cite{broadbent_universal_2009}:
the step at $(i,j)$ acts on wire $i$ between columns $j$ and $j+1$, and the
vertical $CZ$ gates of column $j+1$ act after it. The forward light cone of
that step in the circuit consists of the steps reachable by
$(i,j)\to(i,j+1)$ and by $(i,j)\to(i\pm1,j+1)$ across a vertical edge at
column $j+1$, which is exactly $F((i,j))$. The reduced state of an output
qubit depends only on the gates in its backward light cone, which proves the
claim.

Monotonicity and subadditivity of \eqref{eq:flowdamage} follow because
$F(\cdot)$ distributes over unions. For $D_{\mathrm{bin}}$, if
$F(S)\cap O^\star=\emptyset$ the decision register is in its ideal state, so
the incorrectness event has probability zero. For $D_{\mathrm{frac}}$, each
single-qubit trace distance is at most one and vanishes outside $F(S)$.
\end{proof}

\begin{algorithm}[t]
\caption{Damage profile from the flow}
\label{alg:damage}
\begin{algorithmic}[1]
\Require $G=(V,E)$, flow $(f,\prec)$, outputs $O$, decision set $O^\star$
\For{$u\in V$ in reverse $\prec$-order}
  \State $\mathrm{succ}(u)\gets\{f(u)\}\cup\mathcal N(f(u))\setminus\{u\}$ \Comment{$\emptyset$ if $u\in O$}
  \State $F(u)\gets\{u\}\cup\bigcup_{x\in\mathrm{succ}(u)}F(x)$
\EndFor
\For{$u\in V$}
  \State $w^{\mathrm{bin}}_u\gets\1[F(u)\cap O^\star\neq\emptyset]$
  \State $w^{\mathrm{frac}}_u\gets|F(u)\cap O|/|O|$
\EndFor
\State \Return $w^{\mathrm{bin}},w^{\mathrm{frac}}$
\end{algorithmic}
\end{algorithm}

With bitsets, Algorithm~\ref{alg:damage} runs in $O(|V|\,|E|)$ word
operations, which is negligible next to the delegated computation. For
$D_{\mathrm{frac}}$, the argument of Lemma~\ref{lem:reduction} and of
Theorem~\ref{thm:general} bounds $\E[L\cdot\1_{\text{accept}}]$ for the output
loss $L\in[0,1]$ in place of $\Pr[\text{accept}\wedge\text{incorrect}]$.

\subsection{Exact simulation}
\label{sec:flow-sim}

Proposition~\ref{prop:influence} gives an upper bound on the set of affected
outputs. To test both its soundness and its tightness, we simulated padded
brickwork patterns exactly with a state-vector code that keeps two columns
alive at a time, applies the attack to each qubit after all its $CZ$ gates and
immediately before its measurement, and adapts angles by \eqref{eq:adapt}.
For each instance we used three sets of random angles, applied $X$, $Y$ and
$Z$ at every position with two to three sampled measurement branches each,
and added $60$ random attacks on two to four positions per angle set. An
output was counted as affected when its reduced density matrix differed from
the ideal one by more than $10^{-8}$ in any entry.

\begin{table}[t]
\centering
\caption{Exact simulation of padded brickwork patterns. ``Violations'' counts
affected outputs outside the flow cone. The last two columns compare the
number of (position, output) pairs in the flow cone with the number observed
to be affected. The ideal runs were deterministic to $1.5\times10^{-15}$.}
\label{tab:sim}
\small
\setlength{\tabcolsep}{4pt}
\begin{tabular}{@{}cccccc@{}}
\toprule
$n$ & $m$ & qubits & violations & pairs in cone & pairs affected\\
\midrule
3 & 5 & 15 & 0 & 23 & 23\\
4 & 5 & 20 & 0 & 36 & 36\\
5 & 5 & 25 & 0 & 41 & 41\\
4 & 13 & 52 & 0 & 148 & 148\\
5 & 13 & 65 & 0 & 189 & 189\\
7 & 13 & 91 & 0 & 287 & 287\\
5 & 21 & 105 & 0 & 385 & 385\\
\bottomrule
\end{tabular}
\end{table}

Table~\ref{tab:sim} shows that the flow cone was never violated and that it
was attained exactly: every (position, output) pair allowed by the cone was
affected by some single-qubit Pauli. In these instances the flow cone is
the exact influence set. The simulation also confirms the
correction rule \eqref{eq:adapt}, since the ideal outputs were independent of
the sampled branches to machine precision.

\subsection{Case study: brickwork states}

\begin{figure*}[!tp]
\centering
\includegraphics[width=0.78\textwidth]{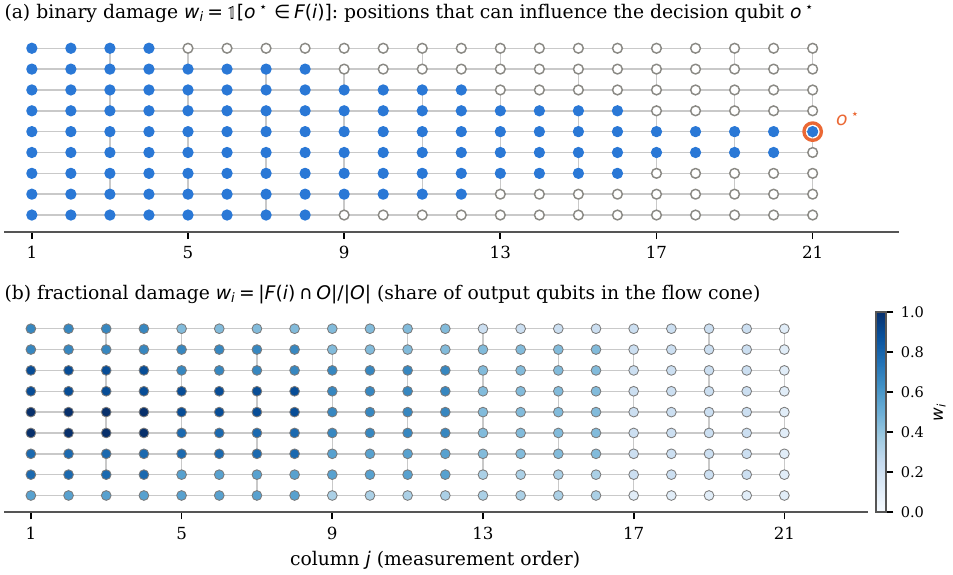}
\caption{Damage profiles of the brickwork state $G_{9\times21}$ from
Algorithm~\ref{alg:damage}. (a) Binary profile for the middle-row output
$o^\star$ (circled): filled vertices lie in the flow cone of $o^\star$,
$117$ of $N=189$. By Table~\ref{tab:sim}, these are the positions at which a
Pauli deviation can affect $o^\star$. (b) Fractional profile: the share of
output qubits inside the flow cone of each vertex.}
\label{fig:brickwork}
\end{figure*}

We apply Algorithm~\ref{alg:damage} to brickwork states $G_{n\times m}$ with
$m\equiv5\pmod8$ columns, following the edge rule of
Ref.~\cite{broadbent_universal_2009}, with the decision qubit $o^\star=(r,m)$
in the middle row. Figure~\ref{fig:brickwork} shows both profiles for $n=9$
and $m=21$. The flow cone of $o^\star$ has a staircase shape and contains
$117$ of the $189$ positions; the fractional profile decays toward later
columns and toward the boundary rows.

\begin{proposition}[Cone size for brickwork patterns]
\label{prop:cone}
Let $m=4h+1$ with $m\equiv5\pmod8$, let $o^\star=(r,m)$ and let
$B=\{u:o^\star\in F(u)\}$. If $h+1\le r\le n-h$, then the rows of $B$ in
column $j<m$ form an interval of $2\lceil(m-j)/4\rceil$ rows, column $m$
contributes $o^\star$ alone, and
\begin{equation}
  |B|=1+4h(h+1)=(2h+1)^2=\Bigl(\frac{m+1}2\Bigr)^2,
  \label{eq:cone-size}
\end{equation}
independently of the width $n$.
\end{proposition}

\begin{proof}
Vertical edges of the brickwork graph occur at every odd column and
alternate, with period eight, between the pairings $\{(2\ell-1,2\ell)\}$ and
$\{(2\ell,2\ell+1)\}$~\cite{broadbent_universal_2009}. Predecessors of
$(i,j)$ in the dependency DAG are $(i,j-1)$, $(i,j-2)$ and the qubits
$(i',j-1)$ such that $(i',j)$ is a vertical neighbor of $(i,j)$. Tracing back
from $o^\star$, column $m-1$ contains $o^\star$'s row and its vertical
partner at column $m$. Moving back, the interval of rows gains one row on one
side at each odd column, and the two pairings alternate, so it gains one row
on each side per block of four columns. The hypothesis on $r$ keeps the
interval away from the boundary of the register. This gives the stated
widths, and
\[
  |B|=1+\sum_{t=1}^{h}4\cdot2t=1+4h(h+1)=(2h+1)^2 .\qedhere
\]
\end{proof}

We verified \eqref{eq:cone-size} with Algorithm~\ref{alg:damage} for
$m=5$, $13$, $21$, $29$ and $37$ at $n=65$, and at the smallest admissible widths
$(n,m)\in\{(7,13),(11,21),(15,29)\}$. For $(n,m)=(7,13)$ the simulation of
Table~\ref{tab:sim} observes all $49$ positions of $B$ affecting $o^\star$.

\begin{figure}[!tb]
\centering
\includegraphics[width=\columnwidth]{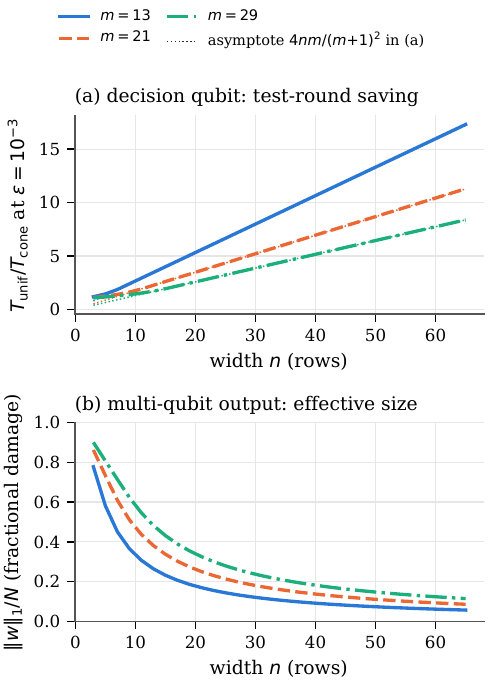}
\caption{(a) Test-round saving $T_{\mathrm{unif}}/T_{\mathrm{cone}}$ for
$\varepsilon=10^{-3}$ against persistent deviations, as a function of the
width of the brickwork state. Dotted lines show the ratio
$N/|B|=4nm/(m+1)^2$ of Proposition~\ref{prop:cone}, which is also the saving
against general servers. (b) Effective size $\|w\|_1/N$ of the fractional
profile.}
\label{fig:brickscaling}
\end{figure}

For the binary profile, $w$ is the indicator of $B$, so $\pw$ is uniform on
$B$ and $\|w\|_1=|B|$. The ratio $N/|B|=4nm/(m+1)^2$ grows linearly in the
width and tends to $4n/m$ for large $m$. By Corollaries~\ref{cor:resources}
and~\ref{cor:resources-general}, it is the factor by which damage-weighted
placement reduces the number of test rounds, against persistent deviations
and against general servers alike. Table~\ref{tab:brickwork} gives exact
values for persistent deviations; the ratio reaches $17.3$ for $n=65$ and
$m=13$ (Figure~\ref{fig:brickscaling}a). Placement optimization is
therefore most useful for wide, shallow computations, where a single
decision qubit depends on a small fraction of the resource state.

For multi-qubit outputs, the fractional profile has effective size
$\|w^{\mathrm{frac}}\|_1$ that is a small fraction of $N$ for wide patterns
(Figure~\ref{fig:brickscaling}b), with $\wmax$ as low as $0.09$. By
Theorem~\ref{thm:robust}, $\pw$ is then within a factor $1.09$ of the robust
single-round optimum. The gain over uniform placement is smaller than for
the binary profile, a factor between $1.4$ and $1.6$ in
Table~\ref{tab:colgen}, because a fractional profile already spreads the
damage over the whole pattern.

\begin{table*}[!tp]
\centering
\caption{Brickwork states, $c=\tfrac12$. $N=nm$; $|B|$ is the size of the flow
cone of the middle-row decision qubit; $T_{\mathrm{unif}}$ and
$T_{\mathrm{cone}}$ are the numbers of test rounds needed for
$\varepsilon\le10^{-3}$ against persistent deviations with uniform placement
and with placement uniform on $B$ (Corollary~\ref{cor:resources}); against
general servers the ratio of the sufficient numbers of
Corollary~\ref{cor:resources-general} is $N/|B|$. The last two columns
describe the fractional profile.}
\label{tab:brickwork}
\small
\begin{tabular}{@{}rrrrrrrrcc@{}}
\toprule
$n$ & $m$ & $N$ & $|B|$ & $T_{\mathrm{unif}}$ & $T_{\mathrm{cone}}$ & saving & $N/|B|$ & $\|w^{\mathrm{frac}}\|_1/N$ & $\wmax^{\mathrm{frac}}$\\
\midrule
 9 & 13 &  117 &  49 &  1613 &  674 & $2.39\times$ & 2.39 & 0.366 & 0.667\\
17 & 13 &  221 &  49 &  3050 &  674 & $4.53\times$ & 4.51 & 0.207 & 0.353\\
33 & 13 &  429 &  49 &  5924 &  674 & $8.79\times$ & 8.76 & 0.110 & 0.182\\
65 & 13 &  845 &  49 & 11671 &  674 & $17.32\times$ & 17.24 & 0.057 & 0.092\\
 9 & 21 &  189 & 117 &  2608 & 1613 & $1.62\times$ & 1.62 & 0.511 & 1.000\\
17 & 21 &  357 & 121 &  4929 & 1669 & $2.95\times$ & 2.95 & 0.303 & 0.588\\
33 & 21 &  693 & 121 &  9571 & 1669 & $5.73\times$ & 5.73 & 0.165 & 0.303\\
65 & 21 & 1365 & 121 & 18855 & 1669 & $11.30\times$ & 11.28 & 0.086 & 0.154\\
 9 & 29 &  261 & 189 &  3603 & 2608 & $1.38\times$ & 1.38 & 0.624 & 1.000\\
17 & 29 &  493 & 225 &  6808 & 3106 & $2.19\times$ & 2.19 & 0.390 & 0.824\\
33 & 29 &  957 & 225 & 13218 & 3106 & $4.26\times$ & 4.25 & 0.217 & 0.424\\
65 & 29 & 1885 & 225 & 26039 & 3106 & $8.38\times$ & 8.38 & 0.115 & 0.215\\
\bottomrule
\end{tabular}
\end{table*}

\section{Test Rounds with Many Traps}
\label{sec:trapsets}

In test-round protocols~\cite{leichtle_verifying_2021} and trappified
schemes~\cite{kapourniotis_unifying_2024}, a single round places traps on
every vertex of an independent set $I$ of the resource graph $G$ and dummies
elsewhere. Leichtle \emph{et al.}\ choose $I$ as a color class of a proper
coloring, with the color drawn uniformly at random. The placement question is then which
distribution $\mu$ over independent sets the client should use. Write
$\pi_i=\Pr_\mu[i\in I]$ for the coverage of vertex $i$.

\begin{lemma}[Detection with trap sets]
\label{lem:trapset}
Under Hypotheses~\ref{ass:twirl}--\ref{ass:damage}, applied independently
to each trap, a round with $I\sim\mu$ satisfies
$\Pr[\text{accept}\wedge\text{incorrect}]\le\max_SD(S)\bigl(1-c\Pr_\mu[I\cap S\neq\emptyset]\bigr)$,
and $\Pr_\mu[I\cap S\ne\emptyset]\ge\max_{i\in S}\pi_i$.
\end{lemma}

\begin{proof}
Conditioned on $I$ with $I\cap S\ne\emptyset$, the traps in $S$ flag
independently with probability at least $c$ each, so no trap flags with
probability at most $1-c$. The rest follows as in Lemma~\ref{lem:reduction}.
\end{proof}

The proof of Theorem~\ref{thm:general}(1) uses only the per-round
non-detection bound, so against general servers it holds for trap-set rounds
with $p(S)$ replaced by the coverage $\Pr_\mu[I\cap S\ne\emptyset]$ in the
definition~\eqref{eq:rho} of $\rho_D$.

For flat damage, the optimal distribution over independent sets was
determined by Kapourniotis \emph{et al.}~\cite[Sec.~6.2.1, Lemmas~6
and~7]{kapourniotis_unifying_2024}: the best worst-case detection
probability of a test round with standard traps is $1/\chi_f(G)$, and an
optimal fractional coloring attains it. Theorem~\ref{thm:fractional}
restates that result in the notation of our game, with the detection
constant $c$; we include the short proof only to make the reduction to the
game explicit. What is new in this section is the damage-weighted version
of the problem, the robust optimality of uniform coloring for brickwork
decision profiles (Corollary~\ref{cor:brick}), and the analysis of rounds
with a limited number of traps in Section~\ref{sec:ktraps}.

\begin{theorem}[Fractional coloring is optimal for flat damage;
recovery of~{\cite[Lemmas~6 and~7]{kapourniotis_unifying_2024}}]
\label{thm:fractional}
Let $D(S)=1$ for every non-empty $S$. The optimal single-round soundness
bound over all distributions on independent sets is exactly
$1-c/\chi_f(G)$, where $\chi_f$ is the fractional chromatic
number~\cite{scheinerman_fractional_1997}. An optimal fractional coloring
attains it. A uniformly chosen color class of a proper $k$-coloring
attains $1-c/k$, and therefore $1-c/\chi(G)$ at best.
\end{theorem}

\begin{proof}
By Lemma~\ref{lem:trapset}, any $\mu$ has soundness bound at most
$1-c\min_i\pi_i$, since every non-empty $S$ contains a vertex. Singleton
attacks show that the bound is at least $1-c\min_i\pi_i$. So the optimum is
$1-c\,t^\ast$, where $t^\ast=\max_\mu\min_i\pi_i$. Rescaling $y=\mu/t$ turns
the linear program for $t^\ast$ into
$\min\{\sum_Iy_I:\sum_{I\ni i}y_I\ge1\ \forall i,\ y\ge0\}$, which is
$\chi_f(G)$ by \eqref{eq:chif}, so
$t^\ast=1/\chi_f(G)$. A uniform color class covers every vertex with
probability $1/k$.
\end{proof}

Since $\omega(G)\le\chi_f(G)\le\chi(G)$, uniform coloring is optimal on
perfect graphs, and in particular on bipartite graphs. The fractional
chromatic number can, however, be much smaller than the chromatic
number~\cite{scheinerman_fractional_1997}. For the five-cycle, which is
also the example of Ref.~\cite[Example~5]{kapourniotis_unifying_2024},
$\chi_f=5/2$ and $\chi=3$. With $c=\tfrac12$ the optimal round achieves $0.8000$, using
the five maximum independent sets uniformly, while a uniform $3$-coloring
achieves $0.8333$. Computing $\chi_f$ is hard in general but polynomial for
perfect graphs~\cite{grotschel_geometric_1988}.

\begin{corollary}[Brickwork test rounds]
\label{cor:brick}
The brickwork state is bipartite. If two adjacent vertices both have
$w_i=1$, then every $\mu$ has soundness bound at least $1-c/2$, and a uniform
$2$-coloring attains $1-c/2$ against every $D\in\cD(w)$. Uniform coloring
is therefore robustly optimal for the binary decision-qubit profile of
Section~\ref{sec:flow} for all $(n,m)$ in Table~\ref{tab:brickwork}.
\end{corollary}

\begin{proof}
Independence gives $\pi_u+\pi_v\le1$ on every edge, so one of the two
adjacent weight-one vertices has coverage at most $\tfrac12$, and a
singleton attack on it pays at least $1-c/2$. A uniform $2$-coloring gives
$\Pr[I\cap S\ne\emptyset]\ge\tfrac12$ for every non-empty $S$, and $D\le1$.
In every instance of Table~\ref{tab:brickwork}, the adjacent vertices
$(\lceil n/2\rceil,1)$ and $(\lceil n/2\rceil,2)$ both have binary weight
one.
\end{proof}

For non-flat damage the optimistic problem becomes a linear program over
coverage vectors in the stable-set polytope $\STAB(G)$:
$\min_{\pi\in\STAB(G)}\max_iw_i(1-c\pi_i)$. For bipartite graphs
$\STAB(G)=\{\pi\ge0:\pi_u+\pi_v\le1\ \forall uv\in E\}$~\cite{schrijver_combinatorial_2003},
so this program is polynomial. For the fractional brickwork profiles with
$(n,m)\in\{(9,13),(17,13),(33,13),(17,21)\}$ it returns exactly the uniform
$2$-coloring value $\wmax(1-c/2)$. On non-bipartite graphs non-uniform
distributions help. On the five-cycle with $w=(1,0.6,0.3,0.3,0.6)$, the
optimistic optimum is $0.5625$ and the robust optimum, an exact linear
program over independent sets and supports, is $0.75$. A uniform
$3$-coloring gives $0.8333$ and uniform maximum independent sets give
$0.8$. As in Sections~\ref{sec:game}--\ref{sec:budget}, the placement
distribution matters most when a round covers only a small fraction of the
damage-relevant positions, as in single-trap constructions, or when $\chi_f$
is noticeably smaller than $\chi$.

\subsection{Rounds with a limited number of traps}
\label{sec:ktraps}

The large savings reported in Sections~\ref{sec:budget} and~\ref{sec:flow}
compare single-trap rounds. Protocols that place a full color class of traps
in every test round~\cite{leichtle_verifying_2021} do not face that
constraint, and for brickwork decision profiles
Corollary~\ref{cor:brick} shows that they gain nothing from optimized
placement. Between these two extremes the number of traps per round may be
limited, for instance when traps are embedded in the same graph as the
computation, as in single-graph constructions, or when the client can
prepare only a limited number of trap states per round. The next
proposition shows that the advantage of computation-dependent placement
persists for every trap budget below half the size of the cone.

\begin{proposition}[Rounds with at most $k$ traps]
\label{prop:ktraps}
Let $G$ be bipartite with proper $2$-coloring $V=V_0\cup V_1$, let
$w=\1_B$ be a binary profile, so that $\cD(w)$ contains only
$D_{\mathrm{bin}}(S)=\1[S\cap B\ne\emptyset]$, and write $B_j=B\cap V_j$.
Consider distributions $\mu$ over independent sets with $|I|\le k$, where
$1\le k\le\min\{|B_0|,|B_1|\}$.
\begin{enumerate}
\item Every such $\mu$ has a position $i\in B$ with $\pi_i\le k/|B|$.
      Its single-round soundness bound is therefore at least $1-ck/|B|$, and
      its damage-to-coverage ratio
      $\rho(\mu)=\max_{S:\,S\cap B\ne\emptyset}1/\Pr_\mu[I\cap S\ne\emptyset]$
      is at least $|B|/k$.
\item Let $\mu_B^k$ choose $j\in\{0,1\}$ with probability $|B_j|/|B|$
      and then a uniformly random $k$-subset of $B_j$. Every $i\in B$ is
      covered with probability exactly $k/|B|$, and $\mu_B^k$ attains the
      single-round bound $1-ck/|B|$, the bound $(1-ck/|B|)^T$ over $T$
      test rounds against persistent deviations, and the general-server bound
      $\min\{1,|B|/(ck(T+1))\}$.
\item The same construction with $V$ in place of $B$, which ignores the
      computation, gives $1-ck/N$, $(1-ck/N)^T$ and
      $\min\{1,N/(ck(T+1))\}$ for $k\le\min\{|V_0|,|V_1|\}$.
\end{enumerate}
In particular, for every admissible $k$ the general-server bounds of the two
constructions differ by the factor $N/|B|$, as for a single trap.
\end{proposition}

\begin{proof}
(1) Since $|I|\le k$, $\sum_{i\in B}\pi_i=\E|I\cap B|\le k$, so some
$i\in B$ has $\pi_i\le k/|B|$. A deviation on $\{i\}$ has damage one and is
detected with probability at most $c\pi_i$, which gives both claims.
(2) Subsets of $B_j\subseteq V_j$ are independent. A position $i\in B_j$ is
covered with probability $(|B_j|/|B|)(k/|B_j|)=k/|B|$. Every $S$ with
$D_{\mathrm{bin}}(S)>0$ meets $B$, so
$\Pr[I\cap S\ne\emptyset]\ge k/|B|$ by Lemma~\ref{lem:trapset}, and the
single-round bound follows from the same lemma. Under
Hypothesis~\ref{ass:rounds} each test round draws its trap set
independently and fails to flag with probability at most $1-ck/|B|$, as in
the proof of Lemma~\ref{lem:reduction}. The general-server bound is
Theorem~\ref{thm:general}(1) with $\rho=|B|/k$.
(3) Replace $B$ by $V$ and use $D\le1$.
\end{proof}

Table~\ref{tab:ktraps} evaluates Proposition~\ref{prop:ktraps} for the
decision qubit of $G_{65\times13}$, for which $|B_0|=25$ and $|B_1|=24$.
The factor between the two constructions stays at $17.2$ against general
servers for every $k\le24$. With $k=24$, placement on the cone needs $25$
persistent test rounds and $4084$ rounds against general servers. A uniform
$2$-coloring, which uses a full color class of about $420$ traps per round,
covers every position with probability $\tfrac12$ and needs $25$ and $4000$
rounds. Placement that depends on the computation thus reaches almost the
same guarantee for this decision qubit with about one seventeenth of the
traps. When the trap budget is unrestricted, the optimized placement and the
full color class coincide in value, in agreement with
Corollary~\ref{cor:brick}.

\begin{table}[!t]
\centering
\caption{Rounds with at most $k$ traps for the decision qubit of
$G_{65\times13}$ ($N=845$, $|B|=49$), $c=\tfrac12$ and
$\varepsilon=10^{-3}$ (Proposition~\ref{prop:ktraps}). $v$ is the
single-round soundness bound, $T_{\mathrm{pers}}$ the number of test rounds
needed against persistent deviations and $T_{\mathrm{gen}}+1$ the
sufficient number of rounds against general servers. ``Cone'' uses
$\mu_B^k$, ``uniform'' the computation-independent construction. The last
row is a uniform $2$-coloring with a full color class per round.}
\label{tab:ktraps}
\small
\setlength{\tabcolsep}{3pt}
\begin{tabular}{@{}rcccccc@{}}
\toprule
 & \multicolumn{3}{c}{cone, $\mu_B^k$} & \multicolumn{3}{c}{uniform}\\
\cmidrule(lr){2-4}\cmidrule(l){5-7}
$k$ & $v$ & $T_{\mathrm{pers}}$ & $T_{\mathrm{gen}}+1$ & $v$ & $T_{\mathrm{pers}}$ & $T_{\mathrm{gen}}+1$\\
\midrule
 1 & 0.9898 & 674 & 98\,000 & 0.99941 & 11\,671 & 1\,690\,000\\
 2 & 0.9796 & 336 & 49\,000 & 0.99882 &  5\,834 &   845\,000\\
 4 & 0.9592 & 166 & 24\,500 & 0.99763 &  2\,916 &   422\,500\\
 8 & 0.9184 &  82 & 12\,250 & 0.99527 &  1\,456 &   211\,250\\
16 & 0.8367 &  39 &  6\,125 & 0.99053 &    727 &   105\,625\\
24 & 0.7551 &  25 &  4\,084 & 0.98580 &    483 &    70\,417\\
\midrule
\multicolumn{4}{@{}l}{full color class ($\approx N/2$ traps)} & 0.7500 & 25 & 4\,000\\
\bottomrule
\end{tabular}
\end{table}

\section{Statistical Certification from Trap Flags}
\label{sec:tomography}

A trap round is a detection event and also a measurement. This section
identifies the quantity that the measurement estimates and the sample cost
of estimating it. The flags are those of the test rounds of
Section~\ref{sec:model-template}. Throughout this section the server is
\emph{stationary}: its effective deviation is drawn independently in each
round from a fixed Pauli mixture $P$. This is the natural model of a noisy
device and of a fixed attack; it is neither implied by nor implies
Hypothesis~\ref{ass:rounds}. Remark~\ref{rem:adaptive} removes the
stationarity requirement.

\subsection{What a trap round estimates}

Let $q_i=\Pr_{E\sim P}[i\in\supp E]$ be the single-qubit marginals of the
twirled deviation. Let $r_i=\Pr[\text{flag}\mid t=i]$ be the \emph{flag
rate} at position $i$, and let $\xi_t\in\{0,1\}$ indicate a flag in test
round $t$.

\begin{proposition}[Flag rates bracket the marginals]
\label{prop:estimator}
Under Hypotheses~\ref{ass:twirl}--\ref{ass:detect},
$c\,q_i\le r_i\le q_i$ for every $i$. If the trap position is drawn from
$p$, then $\E[\xi_t]=\sum_ip_ir_i$.
\end{proposition}

\begin{proof}
Conditioned on $E$ and $t=i$, the trap flags with probability in $[c,1]$ if
$i\in\supp E$ and with probability $0$ otherwise. Averaging over $E\sim P$
gives the bracket. Averaging over $t\sim p$ gives the second claim.
\end{proof}

The client therefore learns the flag rates $r$, which determine the
marginals $q$ only up to the factor $c$. With $c=\tfrac12$, one flag rate is
consistent with marginals differing by a factor of two, depending on which
Pauli types the server uses. An unbiased estimator of $q$ itself would
require the detection probability to equal $c$ exactly for every Pauli type,
which Hypothesis~\ref{ass:detect} does not provide.

\subsection{Certifying the damage functional}

The quantity that matters for soundness is the aggregate damage
$\bar\Lambda=\sum_iw_iq_i/\|w\|_1$, which is bracketed by
$\bar R=\sum_iw_ir_i/\|w\|_1$ via $\bar R\le\bar\Lambda\le\bar R/c$. For a
placement $p$ with $\supp w\subseteq\supp p$, the importance-weighted
statistic $Y_t=w_{i_t}\xi_t/(\|w\|_1\,p_{i_t})$ is unbiased for $\bar R$ and
takes values in $[0,\beta_p]$, where
$\beta_p=\max_{i:w_i>0}w_i/(\|w\|_1p_i)$. In the notation of
\eqref{eq:rho}, $\beta_p=\rho_{\Dmax}(p)/\|w\|_1$, so the range of the
statistic and the general-server soundness bound of
Theorem~\ref{thm:general} are controlled by the same quantity.

\begin{proposition}[Range-optimal certification]
\label{prop:certify}
$\beta_p\ge1$, with equality if and only if $p=\pw$. With $T$ test rounds,
$\Pr[|\bar Y-\bar R|\ge\epsilon]\le2\exp(-2T\epsilon^2/\beta_p^2)$. For
$p=\pw$, $T\ge\ln(2/\delta)/(2\epsilon^2)$ rounds therefore suffice,
independently of $N$.
\end{proposition}

\begin{proof}
$\beta_p\ge\beta_p\sum_{i:w_i>0}p_i\ge\sum_iw_i/\|w\|_1=1$, with equality
if and only if $p_i\propto w_i$. The tail bound is Hoeffding's
inequality~\cite{hoeffding_probability_1963}.
\end{proof}

\begin{remark}[Adaptive servers]
\label{rem:adaptive}
If the server adapts its deviation to past rounds, the terms $Y_t$ are no
longer independent. The centred sums $\sum_t(Y_t-\E[Y_t\mid\mathcal F_{t-1}])$
still form a martingale with increments in intervals of length $\beta_p$.
The Azuma--Hoeffding inequality~\cite{azuma_weighted_1967} then gives the
same tail bound for $\bar Y$ around the time-averaged conditional rate
$T^{-1}\sum_t\bar R_t$.
\end{remark}

\subsection{Localization is necessarily expensive}

\begin{theorem}[Lower bounds]
\label{thm:lower}
Let $\epsilon\le1/8$ and $\delta<1/4$, and consider flag-rate vectors in
$[\tfrac12,\tfrac12+2\epsilon]^N$.
\begin{enumerate}
\item Any procedure, possibly with adaptively chosen trap positions, that
      outputs $\hat r$ with $\max_i|\hat r_i-r_i|<\epsilon$ with probability
      at least $1-\delta$ for every such vector uses
      $T\ge\dfrac{N\ln(1/4\delta)}{16\epsilon^2}$ rounds.
\item Any procedure that estimates the average rate $\bar R$ to accuracy
      $\epsilon$ with the same confidence uses
      $T\ge\ln(1/4\delta)/(16\epsilon^2)$ rounds.
\item Uniform placement with empirical rates estimates all $N$ rates to
      accuracy $\epsilon$ with confidence $1-\delta$ once
      $T\ge\max\{N\epsilon^{-2}\ln(4N/\delta),\,8N\ln(2N/\delta)\}$.
\end{enumerate}
Localizing the deviation therefore costs $\Theta(N)$ times more than
certifying it, up to a factor $\log N$.
\end{theorem}

\begin{proof}
(1) Let $P_0$ be the law of the transcript when every $r_i=\tfrac12$, and let
$P_j$ be the law when $r_j=\tfrac12+2\epsilon$ and all other rates equal
$\tfrac12$. Let $n_j$ be the number of rounds with the trap at $j$. Since
$\sum_j\E_0[n_j]=T$, some $j$ has $\E_0[n_j]\le T/N$. By the divergence
decomposition for adaptively sampled
observations~\cite[Lemma~15.1]{lattimore_bandit_2020},
$\mathrm{KL}(P_0\|P_j)=\E_0[n_j]\,\mathrm{kl}(\tfrac12\|\tfrac12+2\epsilon)$.
Here $\mathrm{kl}(\tfrac12\|\tfrac12+x)=-\tfrac12\ln(1-4x^2)\le2x^2/(1-4x^2)$,
which is at most $16\epsilon^2$ for $x=2\epsilon\le\tfrac14$. The test
``$\hat r_j\ge\tfrac12+\epsilon$'' has error probability at most $\delta$
under both hypotheses. The Bretagnolle--Huber
inequality~\cite[Lemma~2.6]{tsybakov_introduction_2009} gives
$2\delta\ge\tfrac12\exp(-\mathrm{KL})$, so
$(T/N)\,16\epsilon^2\ge\ln(1/4\delta)$.
(2) The same argument applies to the hypotheses ``all rates $\tfrac12$''
and ``all rates $\tfrac12+2\epsilon$'', with $\mathrm{KL}\le16T\epsilon^2$.
(3) A Chernoff bound gives $n_i\ge T/(2N)$ for all $i$ with probability at
least $1-\delta/2$ when $T\ge8N\ln(2N/\delta)$. Conditioned on the counts,
the flags are independent, and Hoeffding's inequality with a union bound
gives accuracy $\epsilon$ with probability at least $1-\delta/2$ when
$T/(2N)\ge\ln(4N/\delta)/(2\epsilon^2)$.
\end{proof}

\begin{figure}[t]
\centering
\includegraphics[width=\columnwidth]{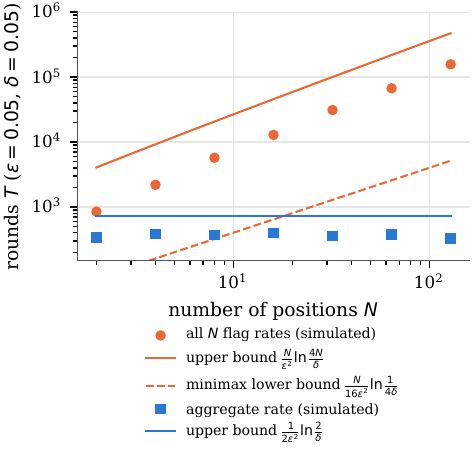}
\caption{Rounds needed to reach accuracy $\epsilon=0.05$ with success
probability $0.95$, estimated from 400 simulated repetitions per point,
with rates drawn uniformly from $[0.3,0.5]$. Squares show certification of
the aggregate rate, which is independent of $N$. Circles show estimation of
all $N$ rates. Lines show the upper bounds of
Proposition~\ref{prop:certify} and Theorem~\ref{thm:lower}(3) and the
minimax lower bound of Theorem~\ref{thm:lower}(1).}
\label{fig:samples}
\end{figure}

Figure~\ref{fig:samples} compares these bounds with simulation. Estimating
the aggregate rate to $\epsilon=0.05$ with 95\% confidence takes between $328$ and $399$ rounds for every $N$ from $2$ to $128$. Estimating all rates
takes $849$ rounds for $N=2$ and $157{,}207$ rounds for $N=128$.

\subsection{Noise-robust certification}

On noisy hardware, traps flag even when the server is honest, which is the
over-sensitivity that motivated noise-robust
protocols~\cite{leichtle_verifying_2021,gheorghiu_robustness_2015}. Suppose
an honest but noisy server has flag rates at most $\gamma_i$, with $\gamma$
known from calibration. Write $\bar\Gamma=\sum_iw_i\gamma_i/\|w\|_1$.

\begin{proposition}[Threshold test, a certification statement]
\label{prop:noise}
Use placement $p$ and accept if and only if $\bar Y\le\bar\Gamma+\tau$.
Then:
(a) an honest noisy server is accepted with probability at least
$1-\exp(-2T\tau^2/\beta_p^2)$;
(b) any server, noisy or malicious, whose effective deviation has
$\bar\Lambda\ge(\bar\Gamma+\tau+\epsilon)/c$ is accepted with probability at
most $\exp(-2T\epsilon^2/\beta_p^2)$.
\end{proposition}

\begin{proof}
(a) $\E[\bar Y]\le\bar\Gamma$, and one-sided Hoeffding applies. (b) By
Proposition~\ref{prop:estimator}, $\bar R\ge c\bar\Lambda\ge\bar\Gamma+\tau+\epsilon$,
and one-sided Hoeffding applies again.
\end{proof}

The test certifies the total weighted rate of non-trivial effective errors,
irrespective of their origin. For a stationary server the flags are
independent Bernoulli variables with parameters $r_{i_t}$, so their law
depends on the deviation only through the flag rates $r$. No test built
from the flags can separate a malicious deviation from honest
noise that produces the same flag rates. The server may hide an attack inside the noise budget, which is why
noise-robust protocols pair such tests with a computation that tolerates a
bounded error rate, and combine the results of several computation
rounds~\cite{leichtle_verifying_2021}. The damage-weighted placement
minimizes $\beta_p$ and hence the number of rounds for given
$(\tau,\epsilon)$. Minimizing the completeness and soundness errors jointly
over $p$ when $\gamma$ is non-uniform is a linear-fractional problem that
we leave open.

Proposition~\ref{prop:noise} is a statement about the aggregate rate
$\bar\Lambda$, not a soundness guarantee in the sense of
Section~\ref{sec:model-template}. With a single computation round the
threshold rule cannot be sound, because it tolerates flags.

\begin{proposition}[The threshold rule is not sound with one computation round]
\label{prop:notsound}
Run the protocol template of Section~\ref{sec:model-template} with
placement $p$, $\supp w\subseteq\supp p$, and the threshold rule of
Proposition~\ref{prop:noise} with tolerance $\tau>0$. Let $S$ be a support
with $w(S)/\|w\|_1\le\tau-\epsilon$ for some $\epsilon\in(0,\tau]$, and
suppose that a deviation on $S$ makes the computation round incorrect with
probability $d_S>0$. Consider a server whose device is noiseless and which
applies this deviation in every round. Then, for every $T$,
\[
  \Pr[\text{accept}\wedge\text{incorrect}]\ge
  d_S\bigl(1-\exp(-2T\epsilon^2/\beta_p^2)\bigr).
\]
\end{proposition}

\begin{proof}
A trap outside $S$ never flags (Hypothesis~\ref{ass:detect}) and a trap in
$S$ flags with probability at most one, so
$\E[Y_t]=\bar R\le w(S)/\|w\|_1\le\tau-\epsilon\le\bar\Gamma+\tau-\epsilon$.
The $Y_t$ are independent and take values in $[0,\beta_p]$, so Hoeffding's
inequality gives $\Pr[\bar Y>\bar\Gamma+\tau]\le\exp(-2T\epsilon^2/\beta_p^2)$.
By Hypothesis~\ref{ass:damage} the computation round is incorrect
independently of the flags, with probability $d_S$.
\end{proof}

The bound does not decay with $T$. For the decision qubit of a brickwork
pattern and $p=\pw$ uniform on the cone $B$, a single position has
$w(S)/\|w\|_1=1/|B|$, which is $0.02$ for $|B|=49$; any tolerance above
this value lets a persistent deviation on one cone position through with
high probability, and such a deviation can change the decision qubit
(Table~\ref{tab:sim}). This is why noise-robust protocols combine the threshold
test with several computation rounds, a majority vote, and computations
that tolerate a bounded error rate~\cite{leichtle_verifying_2021}. In this
paper the threshold rule is therefore used only as a statistical
certification procedure, and the soundness statements of
Theorem~\ref{thm:protocol} refer to the rule that aborts on any flag.

\section{Detection Compared with Reconstruction}
\label{sec:tension}

The game of Section~\ref{sec:game} and the certification problem of
Section~\ref{sec:tomography} each prescribe a placement. This section
compares them.

\begin{proposition}[When the optimistic and damage-weighted placements coincide]
\label{prop:tension}
Let $w\neq0$ and let $P=\{i:w_i>0\}$. The water-filling placement $p^\star$
of Theorem~\ref{thm:minimax} equals $\pw$ if and only if either $w$ is
constant on $P$, or $c=1$ and $|P|=2$.
\end{proposition}

\begin{proof}
If $w$ is constant on $P$, the argument of Corollary~\ref{cor:fk} restricted
to $P$ gives $p^\star=\pw$, both uniform on $P$. If $c=1$ and $P=\{1,2\}$
then $v_2=w_1w_2/(w_1+w_2)<w_2$, so $k^\ast=2$, and substitution gives
$p^\star_i=w_i/(w_1+w_2)$.

Conversely, suppose $p^\star=\pw$. Every $i\in P$ has $\pw_i>0$, so the
active set is $P$ and $(1-\vmax/w_i)/c=w_i/W$ for all $i\in P$, where
$W=\|w\|_1$. Hence each $w_i$ is a root of $x^2-(W/c)x+\vmax W/c=0$. If $w$
takes two distinct values $a>b$ on $P$, Vieta's formulas give $a+b=W/c$.
Since $a+b\le W$, with equality only when $|P|=2$, we obtain $c\ge1$, hence
$c=1$ and $|P|=2$.
\end{proof}

Proposition~\ref{prop:tension} shows that under the optimistic model the two
objectives prescribe different placements except in degenerate cases. In that model the positions outside the active set are never probed, so
their flag rates remain unconstrained by the data the client collects.

In the robust regime the conflict largely disappears. The same placement
$\pw$ is within a factor $1+\wmax$ of the optimal robust detection margin
(Theorem~\ref{thm:robust}) and is the unique range-optimal design for
certifying aggregate damage (Proposition~\ref{prop:certify}). Against
general servers it is exactly optimal for every damage function in $\cD(w)$
(Theorem~\ref{thm:general}), and the range of the certification statistic
and the general-server bound are governed by the same quantity,
$\beta_p=\rho_{\Dmax}(p)/\|w\|_1$. In the
large-budget regime of Theorem~\ref{thm:collapse} the robustly optimal
placement covers every position with $w_i>0$, so every flag rate becomes
estimable. What remains is a quantitative gap at intermediate budgets, where
the robust optimum and $\pw$ differ by a few percent
(Tables~\ref{tab:budget} and~\ref{tab:colgen}), and a qualitative gap in
localization: by Theorem~\ref{thm:lower} no placement avoids the factor
$\Omega(N)$ that separates learning all flag rates from certifying their
weighted average.

\section{An Assembled Protocol}
\label{sec:protocol}

The components of the previous sections combine into a protocol that a
client can run. Algorithm~\ref{alg:protocol} states it in the template of
Section~\ref{sec:model-template}, with single-trap test rounds; the
trap-set variant of Section~\ref{sec:trapsets} replaces the sampling step by
a draw from a distribution over independent sets, for instance
$\mu_B^k$ of Proposition~\ref{prop:ktraps} when at most $k$ traps are
allowed per round. Theorem~\ref{thm:protocol} gives
its soundness, completeness and cost. The protocol has two decision
modes. In \emph{verification mode} the client aborts whenever any test round
flags; Theorem~\ref{thm:protocol} applies to this mode. In
\emph{certification mode}, intended for noisy devices, the client tolerates
flags up to a threshold; this mode certifies the aggregate damage rate
(Proposition~\ref{prop:noise}) but carries no soundness guarantee with a
single computation round (Proposition~\ref{prop:notsound}).

\begin{algorithm}[t]
\caption{Damage-weighted trap verification}
\label{alg:protocol}
\begin{algorithmic}[1]
\Require pattern $(G,I,O,\phi)$ with flow, decision set $O^\star$, constant
$c$, target soundness $\varepsilon$, regime (persistent or general),
decision mode (verification or certification), optional noise bounds
$\gamma$ and tolerance $\tau$
\Statex \emph{Offline, classical}
\State $\hat w\gets$ damage profile from Algorithm~\ref{alg:damage},
       or any over-approximation $\hat w\ge w$
\State choose the placement: by default $p\gets p^{\hat w}$, for every
       $T$; only in the persistent regime and once $\vmax(T)\le\wmin/2$,
       optionally $p^\star(T)$ of Proposition~\ref{prop:multitrap},
       which is then robustly optimal by Theorem~\ref{thm:collapse}
\State choose $T$: general regime, $T+1\gets\lceil\|\hat w\|_1/(c\,\varepsilon)\rceil$;
       persistent regime, the least $T$ with
       $\max\{(1-c/\|\hat w\|_1)^T,\|\hat w\|_1/(c(T+1))\}\le\varepsilon$, or
       Corollary~\ref{cor:resources} for a binary profile
\Statex \emph{Online, $T+1$ blind rounds on $G$}
\State draw the computation round $\kappa\in[T+1]$ uniformly and secretly
\For{$r=1,\dots,T+1$}
  \If{$r=\kappa$} run the pattern blindly
  \Else
    \State sample $i_r\sim p$ secretly; prepare a trap at $i_r$ and dummies
           elsewhere
    \State record the flag $\xi_r\in\{0,1\}$
  \EndIf
\EndFor
\Statex \emph{Decision, after the last round}
\State verification mode (Theorem~\ref{thm:protocol}): accept if and only if $\sum_{r\ne\kappa}\xi_r=0$
\State certification mode (Proposition~\ref{prop:noise}): accept if and only if
       $\bar Y\le\bar\Gamma+\tau$, with $Y_r$ as in
       Proposition~\ref{prop:certify}; this certifies $\bar\Lambda$ but is
       not a soundness guarantee (Proposition~\ref{prop:notsound})
\State on acceptance output the result of round $\kappa$, otherwise abort
\end{algorithmic}
\end{algorithm}

\begin{theorem}[Protocol guarantees, verification mode]
\label{thm:protocol}
Assume Hypotheses~\ref{ass:twirl}--\ref{ass:damage} with constant $c$, and
let $\hat w\ge w$ be the profile used by Algorithm~\ref{alg:protocol}, with
$\hat W=\|\hat w\|_1\ge\max\{1,2c\}$ and $p=p^{\hat w}$. Let $D\in\cD(w)$.
Statements (1)--(3) concern verification mode, in which the client
aborts whenever any test round flags.
\begin{enumerate}
\item \emph{(Soundness, general servers.)} For every server strategy,
      \begin{equation}
        \Pr[\text{accept}\wedge\text{incorrect}]\le
        \min\Bigl\{1,\;\frac{\hat W}{c\,(T+1)}\Bigr\},
        \label{eq:protocol-general}
      \end{equation}
      and no placement gives a smaller right-hand side
      for the profile $\hat w$ (Theorem~\ref{thm:general}).
\item \emph{(Soundness, persistent deviations.)} If in addition
      Hypothesis~\ref{ass:rounds} holds,
      \begin{equation}
        \begin{aligned}
        &\Pr[\text{accept}\wedge\text{incorrect}]\\
        &\qquad\le\max\Bigl\{\bigl(1-c/\hat W\bigr)^{T},\;\frac{\hat W}{c(T+1)}\Bigr\}.
        \end{aligned}
        \label{eq:protocol-sound}
      \end{equation}
      If the profile is binary with support $B$, the bound is
      $(1-c/|B|)^{T}$. If $\vmax(T)\le\wmin/2$ and $p=p^\star(T)$, the bound
      is $\vmax(T)$.
\item \emph{(Completeness.)} In the noiseless model an honest server is
      accepted with probability $1$. If honest noise makes every trap
      flag with probability at least $\gamma_{\min}>0$, verification mode
      aborts an honest server with probability at least
      $1-(1-\gamma_{\min})^T$; certification mode should then be used, with
      the guarantees of Propositions~\ref{prop:noise}
      and~\ref{prop:notsound} instead of (1) and (2).
\item \emph{(Over-estimated profiles.)} The bounds in (1) and (2) are
      conservative: they hold for the true profile $w$ whenever
      $\hat w\ge w$. Optimality holds for $\hat w$, not necessarily for
      $w$. For every $D\in\cD(w)$,
      $\rho_D(p^{\hat w})\le\|\hat w\|_1=\frac{\|\hat w\|_1}{\|w\|_1}\,
      \min_q\rho_D(q)$, so the general-server bound is within the factor
      $\|\hat w\|_1/\|w\|_1$ of the best bound for the true profile, and
      $p^{\hat w}$ is the optimal placement for $w$ if and only if $\hat w$
      is proportional to $w$.
\item \emph{(Classical cost.)} The profile costs $O(|V|\,|E|)$ word
      operations (Algorithm~\ref{alg:damage}); the placement costs $O(N)$
      for $p^{\hat w}$, and $p^\star(T)$ costs one root finding
      (Proposition~\ref{prop:multitrap}).
\end{enumerate}
\end{theorem}

\begin{proof}
(1) Theorem~\ref{thm:general}(1) and (2) applied to $\hat w$, since every
$D\in\cD(w)$ satisfies $D\le\Dadd^{w}\le\Dadd^{\hat w}$ and
$\Dadd^{\hat w}\in\cD(\hat w)$.
(2) Combine Lemma~\ref{lem:reduction} with Proposition~\ref{prop:pwT},
applied to $\hat w$; the guarantee transfers to $w$ by the same domination.
For a binary profile, every $S$ that meets $B$ has $p(S)\ge1/|B|$ and every
other $S$ has $D(S)=0$. The collapse case is Theorem~\ref{thm:collapse}.
(3) In the noiseless model a trap outside the support of the deviation does
not flag by Hypothesis~\ref{ass:detect}, and an honest server applies no
deviation. Under honest noise each of the $T$ independent test rounds
flags with probability at least $\gamma_{\min}$.
(4) The first inequality was shown in (1). By
Theorem~\ref{thm:general}(2), $\min_q\rho_D(q)=\|w\|_1$ for every
$D\in\cD(w)$, and the minimum is attained only by $\pw$. Since
$p^{\hat w}=\pw$ exactly when $\hat w$ is proportional to $w$, the claim
follows.
(5) Algorithm~\ref{alg:damage} performs one reverse pass with bitset unions;
$p^{\hat w}$ requires normalizing $\hat w$; $\Phi$ in
Proposition~\ref{prop:multitrap} is monotone, so its root is found by
bisection.
\end{proof}

\paragraph{Worked resource count} Consider the brickwork pattern
$G_{65\times13}$, so $N=845$ positions, with the decision qubit in the
middle row and $c=\tfrac12$. Proposition~\ref{prop:cone} gives $|B|=49$. For
$\varepsilon=10^{-3}$ and persistent deviations, the binary case of
\eqref{eq:protocol-sound} (Corollary~\ref{cor:resources}) requires $T=674$
test rounds with traps
placed uniformly on $B$, against $T=11\,671$ with traps placed uniformly
over all positions, a reduction by a factor $17.3$. Against general servers,
the bound \eqref{eq:protocol-general} of Theorem~\ref{thm:protocol}
requires $T+1\ge98\,000$ rounds on $B$
against $1\,690\,000$ for uniform placement, a factor $17.2$. For the
fractional profile of the same pattern, $\|w\|_1=48.14$ and $\wmax=0.092$.
A single test round certifies $1-c/\|w\|_1=0.9896$, and no placement
improves this detection margin by more than the factor $1+\wmax=1.09$
(Theorem~\ref{thm:robust}). Against general servers $\pw$ needs
$T+1\ge9.63\times10^4$, and uniform placement, for which
$\rho_{\Dadd}=N\wmax=78$, needs $1.56\times10^5$. The gain is smaller than
for the binary profile because the fractional profile already spreads
damage over the whole pattern. All factors in this paragraph compare
single-trap rounds. With $k$ traps per round, Table~\ref{tab:ktraps} shows
that the factor stays at $17.2$ against general servers for every
$k\le24$, and that a uniform $2$-coloring with full color classes reaches
almost the same guarantee as $24$ traps on the cone. For the brickwork profiles of
Section~\ref{sec:flow} the profile computed by Algorithm~\ref{alg:damage} is
exact, so $\hat w=w$ and the optimality statements of
Theorem~\ref{thm:protocol} apply to the true profile.

\begin{remark}[Use inside majority-vote protocols]
\label{rem:integration}
In the protocols of Refs.~\cite{leichtle_verifying_2021,kapourniotis_unifying_2024},
several computation rounds are interleaved with test rounds, the client
accepts when the fraction of failed test rounds is below a threshold, and
the output is a majority vote over the computation rounds. The quantity that
enters the analysis is again the probability that a test round detects a
given deviation, which Sections~\ref{sec:robust} and~\ref{sec:trapsets}
change and nothing else. What does not carry over directly is the argument
that converts this probability into a bound on the majority. For
non-uniform placement the per-round detection probability depends on the
support of the deviation, and a direct adaptation would set the threshold
against the smallest detection probability among damaging supports,
$c\min_{i:w_i>0}p_i$, which equals $c\,\wmin/\|w\|_1$ for $\pw$. Whether a
damage-weighted analysis of the majority vote does better is open.
Re-deriving the composable statements of
Ref.~\cite{leichtle_verifying_2021} with this quantity is stated as an open
problem in Section~\ref{sec:discussion}.
\end{remark}

\section{Discussion and Limitations}
\label{sec:discussion}

\subsection{Evidence for the working hypotheses}

Hypothesis~W1 of Section~\ref{sec:hypotheses} asserted that damage profiles
are strongly non-uniform and cheaply computable. Section~\ref{sec:flow}
computes them from the flow in $O(|V||E|)$ word operations, and
Proposition~\ref{prop:cone} shows that for brickwork patterns with
$m=4h+1$ columns only $(2h+1)^2$ of the $nm$ positions can influence a
decision qubit away from the boundary rows. For $n=65$ and $m=13$ this is
$49$ of $845$ positions, or $5.8$ percent of the register. In exact
simulations of blind brickwork patterns with up to $105$ qubits, every
position in the flow cone did change the output under some Pauli deviation
and no position outside it did (Table~\ref{tab:sim}), so the cones are not
loose in these instances. The fractional profiles of
Table~\ref{tab:brickwork} have effective size $\|w\|_1/N$ between $0.057$
and $0.624$. W1 is supported for this family of patterns. Patterns
with gflow rather than flow were not examined, and for them the cone of
Proposition~\ref{prop:influence} would have to be replaced by the
correction sets of the gflow.

Hypothesis~W2 asserted that damage-weighted sampling improves the guarantee
by a factor that grows with non-uniformity, and that it is close to optimal.
For a single round, Theorem~\ref{thm:robust} replaces $1-c/N$ by
$1-c/\|w\|_1$ and bounds the remaining gap by $1+\wmax\le2$;
Table~\ref{tab:colgen} shows the realized gap to be at most $2.4$ percent
for brickwork fractional profiles. The gain over uniform placement is the
factor $N/|B|$, up to $17$ in Table~\ref{tab:brickwork}, for binary profiles,
and a factor between $1.4$ and $1.6$ for the fractional profiles, whose
damage is already spread over the pattern. Against general servers the
damage-weighted placement is exactly optimal (Theorem~\ref{thm:general}).
These gains are relative to computation-independent placement with the
same number of traps per round. They persist for every trap budget below
half the cone (Proposition~\ref{prop:ktraps}), but they disappear when each
test round may use a full color class: for brickwork decision profiles
uniform coloring is then already robustly optimal
(Corollary~\ref{cor:brick}). W2 is supported in these regimes. It fails for persistent deviations at
large budgets, where Theorem~\ref{thm:collapse} shows that the
water-filling placement is robustly optimal and $\pw$ is eventually worse
than uniform placement (Table~\ref{tab:budget}).

Hypothesis~W3 asserted that guarantees tied to one combination rule are
unsafe. Proposition~\ref{prop:optimism} shows that the optimistic optimum
can have robust soundness error $1$, and the water-filling placement had a
strictly worse robust guarantee than $\pw$ in $333$ of $400$ random
profiles. W3 is supported.

\subsection{Scope of the modeling hypotheses}

Lemma~\ref{lem:reduction} and Theorem~\ref{thm:general} rest on
Hypotheses~\ref{ass:twirl}--\ref{ass:damage}, whose status is discussed in
Section~\ref{sec:model-status}. For flat damage the per-round bound
coincides numerically with the single-trap theorems of FK
(Remark~\ref{rem:calibration}). FK analyze a single-graph construction, so
the coincidence motivates the values of $c$ used in the examples but does
not validate the hypotheses for any particular trap design. For a given
protocol, the constant $c$ and the independence properties of
Hypotheses~\ref{ass:detect} and~\ref{ass:damage} have to be taken from that
protocol's own soundness proof. The results of this paper concern the
placement layer above such proofs and do not replace them.

\subsection{Encoded computations and threshold damage}

FK and later protocols amplify soundness by encoding the computation, or by
combining the results of several rounds, so that the output fails only when
at least $d$ positions, or $d$ rounds, are corrupted. The corresponding
damage functions are threshold-like and supermodular, so they lie outside
the class $\cD(w)$ of Definition~\ref{def:damage}, in which $D(\{i\})=w_i$
already accounts for the harm of a single deviation. The placement problem
remains a finite zero-sum game whose value is a linear program over
supports, and the duality of Theorem~\ref{thm:duality} holds verbatim with
$\bar D_S$ replaced by the threshold damage, since its proof uses only
bilinearity. What changes is the structure of the optimum: covers are
replaced by sets of size at least $d$, and the packing argument of
Lemma~\ref{lem:packing} no longer applies. Solving the placement problem for
threshold damages, and combining it with the encoding overhead, is the most
natural extension of this work.

\subsection{Majority-vote protocols}

With one computation round the soundness error against general servers is
$\Theta(\|w\|_1/(cT))$ (Theorem~\ref{thm:general}), and on noisy
devices the threshold rule that tolerates flags is not sound at all
(Proposition~\ref{prop:notsound}). Exponentially small
error against general servers requires several computation rounds, a
failure threshold on the test rounds and a majority vote over the
computation rounds, as in LMKO~\cite{leichtle_verifying_2021}. Placement
enters those protocols through the per-round detection probability of a
damaging deviation, which becomes support-dependent once the placement is
not uniform (Remark~\ref{rem:integration}). We conjecture that the
LMKO analysis goes through with the uniform detection probability replaced
by $c\min_{S:D(S)>0}p(S)$, and that a sharper, damage-weighted analysis of
the majority exists; neither statement is proved here. Combining such an
analysis with the round-level optimization of
Ref.~\cite{vandam_optimizing_2026} is the main open problem left by this
work.

\subsection{Adaptive deviations and composability}

Theorem~\ref{thm:general} covers servers whose deviation in each round
depends arbitrarily on their view of the previous rounds, because under
Hypothesis~\ref{ass:blind} that view is independent of the computation-round
index and of the trap positions. The exponential statements of
Section~\ref{sec:budget-persistent} and Theorem~\ref{thm:protocol}(2) use
Hypothesis~\ref{ass:rounds} and do not cover such servers. The estimation
results extend to adaptive servers through the martingale argument of
Remark~\ref{rem:adaptive}. Establishing composable security in the sense of
abstract cryptography~\cite{dunjko_composable_2014} for a non-uniform trap
distribution, along the lines of
Refs.~\cite{leichtle_verifying_2021,kapourniotis_unifying_2024}, is open,
and we make no formal composability claim.

\subsection{Beyond the Pauli reduction}

Hypothesis~\ref{ass:twirl} reduces deviations to Pauli mixtures. Server
strategies that are coherent across supports are not treated
separately, and neither are strategies that correlate a deviation with the
classical transcript in ways the twirl does not remove. The placement
question for such strategies, and for servers with their own costs, which
makes the game general-sum and gives commitment real
value~\cite{korzhyk_stackelberg_2011,morimae_rational_2020}, is left open.

\subsection{Experimental relevance}

Trap-based and related verification has been demonstrated on
photonic~\cite{barz_experimental_2013}, ion and photon~\cite{drmota_verifiable_2024}
and trapped-ion measurement-based~\cite{ringbauer_verifiable_2025}
platforms, at small sizes. The analysis identifies the regimes in which optimized placement is
worthwhile: constructions with few traps per instance relative to the number
of damage-relevant positions, wide and shallow patterns
(Table~\ref{tab:brickwork}), rounds whose number of traps is limited
(Table~\ref{tab:ktraps}), and resource graphs with $\chi_f<\chi$, the
last case being the one already treated in
Ref.~\cite{kapourniotis_unifying_2024}.
Corollary~\ref{cor:brick} shows the opposite case: uniform
coloring is already optimal for brickwork test rounds with full color
classes, so no gain should be expected there. Deciding which regime a given experiment is in requires only
the damage profile, which Algorithm~\ref{alg:damage} produces from the
pattern.

\section{Conclusion}
\label{sec:conclusion}

The best way to place traps in verifiable blind quantum computation depends
on how damage from several corrupted positions combines, which the client
cannot know exactly. We formulated placement as a zero-sum game in the
hidden test-round setting and solved it uniformly over every monotone
subadditive combination rule consistent with a given damage profile.

The soundness guarantees hold for the rule that aborts on any flagged
trap; the noise-tolerant threshold rule is a statistical certification
procedure, and with a single computation round it is not sound.
For a single round, the optimistic extreme of that family has an elementary
water-filling solution that a server accumulating damage defeats with
certainty, so the optimistic value is a limit on what placement can achieve
rather than a guarantee. The conservative extreme admits an exact dual
characterization in terms of damage loads, is attained to within a factor
$1+\wmax\le2$ in detection margin by the damage-weighted placement
$p\propto w$, which certifies $1-c/\|w\|_1$, and can be computed for
hundreds of positions by column generation. Over many rounds the picture
depends on how the server behaves. Against persistent deviations the error decays
exponentially and the two extremes coincide once $\vmax(T)\le\wmin/2$.
Against general servers, which may deviate in only a few rounds, the error
of a protocol with one computation round is $\Theta(\|w\|_1/(cT))$ for the
damage-weighted placement, and that placement is exactly optimal for every
combination rule in the family.

Damage profiles follow from the flow of the measurement pattern in
polynomial time. The flow cones are sound, and in exact simulations of blind
brickwork patterns they coincide with the observed influence sets. For
brickwork patterns with $m=4h+1$ columns the cone of a decision qubit has
$(2h+1)^2$ positions independently of the width, so the number of test
rounds falls by a factor close to $4nm/(m+1)^2$ in both regimes, relative
to uniform single-trap placement, and by the same factor for any number of
traps per round up to half the cone. For flat
damage and test rounds that carry many traps, the optimal soundness per
round is $1-c/\chi_f(G)$, as shown by Kapourniotis \emph{et
al.}~\cite{kapourniotis_unifying_2024}, so the uniform coloring used by existing
protocols is optimal on bipartite resource states such as the brickwork
state. Finally, trap statistics estimate flag rates that bracket the
marginals of the twirled deviation. Certifying the aggregate damage costs a number of rounds independent of the number of positions, and the damage-weighted placement is the unique range-optimal design for that task, whereas localizing the damage costs a factor $\Omega(N)$ more for any strategy. Extending the placement analysis to majority-vote protocols with several computation rounds is the main open problem.

\bibliographystyle{IEEEtranDOI}
\bibliography{trapgame}

\begin{thebibliography}{10}
\providecommand{\url}[1]{#1}
\providecommand{\bibdoi}[1]{\url{#1}}
\csname url@samestyle\endcsname
\providecommand{\newblock}{\relax}
\providecommand{\bibinfo}[2]{#2}
\providecommand{\BIBentrySTDinterwordspacing}{\spaceskip=0pt\relax}
\providecommand{\BIBentryALTinterwordstretchfactor}{4}
\providecommand{\BIBentryALTinterwordspacing}{\spaceskip=\fontdimen2\font plus
\BIBentryALTinterwordstretchfactor\fontdimen3\font minus
  \fontdimen4\font\relax}
\providecommand{\BIBforeignlanguage}[2]{{%
\expandafter\ifx\csname l@#1\endcsname\relax
\typeout{** WARNING: IEEEtran.bst: No hyphenation pattern has been}%
\typeout{** loaded for the language `#1'. Using the pattern for}%
\typeout{** the default language instead.}%
\else
\language=\csname l@#1\endcsname
\fi
#2}}
\providecommand{\BIBdecl}{\relax}
\BIBdecl

\bibitem{childs_secure_2005}
A.~M. Childs, ``Secure assisted quantum computation,'' \emph{Quantum Inf.
  Comput.}, vol.~5, no.~6, pp. 456--466, 2005, arXiv:quant-ph/0111046, doi:
  \bibdoi{10.26421/QIC5.6-4}.

\bibitem{broadbent_universal_2009}
A.~Broadbent, J.~Fitzsimons, and E.~Kashefi, ``Universal blind quantum
  computation,'' in \emph{2009 50th {Annual} {IEEE} {Symposium} on
  {Foundations} of {Computer} {Science}}.\hskip 1em plus 0.5em minus
  0.4em\relax IEEE, 2009, pp. 517--526, arXiv:0807.4154, doi:
  \bibdoi{10.1109/FOCS.2009.36}.

\bibitem{barz_demonstration_2012}
S.~Barz, E.~Kashefi, A.~Broadbent, J.~F. Fitzsimons, A.~Zeilinger, and
  P.~Walther, ``Demonstration of blind quantum computing,'' \emph{Science},
  vol. 335, pp. 303--308, 2012, doi: \bibdoi{10.1126/science.1214707}.

\bibitem{fitzsimons_private_2017}
J.~F. Fitzsimons, ``Private quantum computation: an introduction to blind
  quantum computing and related protocols,'' \emph{npj Quantum Inf.}, vol.~3,
  p.~23, 2017, doi: \bibdoi{10.1038/s41534-017-0025-3}.

\bibitem{gheorghiu_verification_2019}
A.~Gheorghiu, T.~Kapourniotis, and E.~Kashefi, ``Verification of {Quantum}
  {Computation}: {An} {Overview} of {Existing} {Approaches},'' \emph{Theory
  Comput. Syst.}, vol.~63, no.~4, pp. 715--808, 2019, doi:
  \bibdoi{10.1007/s00224-018-9872-3}.

\bibitem{aharonov_interactive_2017}
D.~Aharonov, M.~Ben-Or, E.~Eban, and U.~Mahadev, ``Interactive proofs for
  quantum computations,'' arXiv:1704.04487, 2017, preliminary version by D.
  Aharonov, M. Ben-Or, and E. Eban in Proceedings of Innovations in Computer
  Science (ICS 2010), pp. 453--469, doi: \bibdoi{10.48550/arXiv.1704.04487}.

\bibitem{reichardt_classical_2013}
B.~W. Reichardt, F.~Unger, and U.~Vazirani, ``Classical command of quantum
  systems,'' \emph{Nature}, vol. 496, pp. 456--460, 2013, doi:
  \bibdoi{10.1038/nature12035}.

\bibitem{mahadev_classical_2018}
U.~Mahadev, ``Classical verification of quantum computations,'' in \emph{2018
  IEEE 59th Annual Symposium on Foundations of Computer Science (FOCS)}.\hskip
  1em plus 0.5em minus 0.4em\relax IEEE, 2018, pp. 259--267, arXiv:1804.01082,
  doi: \bibdoi{10.1109/FOCS.2018.00033}.

\bibitem{fitzsimons_post_2018}
J.~F. Fitzsimons, M.~Hajdu{\v{s}}ek, and T.~Morimae, ``Post hoc {Verification}
  of {Quantum} {Computation},'' \emph{Phys. Rev. Lett.}, vol. 120, no.~4, p.
  040501, 2018, doi: \bibdoi{10.1103/PhysRevLett.120.040501}.

\bibitem{hayashi_verifiable_2015}
M.~Hayashi and T.~Morimae, ``Verifiable {Measurement}-{Only} {Blind} {Quantum}
  {Computing} with {Stabilizer} {Testing},'' \emph{Phys. Rev. Lett.}, vol. 115,
  no.~22, p. 220502, 2015, doi: \bibdoi{10.1103/PhysRevLett.115.220502}.

\bibitem{takeuchi_resource_2019}
Y.~Takeuchi, A.~Mantri, T.~Morimae, A.~Mizutani, and J.~F. Fitzsimons,
  ``Resource-efficient verification of quantum computing using {Serfling}'s
  bound,'' \emph{npj Quantum Inf.}, vol.~5, p.~27, 2019, arXiv:1806.09138, doi:
  \bibdoi{10.1038/s41534-019-0142-2}.

\bibitem{barz_experimental_2013}
S.~Barz, J.~F. Fitzsimons, E.~Kashefi, and P.~Walther, ``Experimental
  verification of quantum computation,'' \emph{Nat. Phys.}, vol.~9, no.~11, pp.
  727--731, 2013, doi: \bibdoi{10.1038/nphys2763}.

\bibitem{drmota_verifiable_2024}
P.~Drmota, D.~P. Nadlinger, D.~Main, B.~C. Nichol, E.~M. Ainley, D.~Leichtle,
  A.~Mantri, E.~Kashefi, R.~Srinivas, G.~Araneda, C.~J. Ballance, and D.~M.
  Lucas, ``Verifiable {Blind} {Quantum} {Computing} with {Trapped} {Ions} and
  {Single} {Photons},'' \emph{Phys. Rev. Lett.}, vol. 132, no.~15, p. 150604,
  2024, doi: \bibdoi{10.1103/PhysRevLett.132.150604}.

\bibitem{fitzsimons_unconditionally_2017}
J.~F. Fitzsimons and E.~Kashefi, ``Unconditionally verifiable blind quantum
  computation,'' \emph{Phys. Rev. A}, vol.~96, no.~1, p. 012303, 2017,
  arXiv:1203.5217, doi: \bibdoi{10.1103/PhysRevA.96.012303}.

\bibitem{kashefi_optimised_2017}
E.~Kashefi and P.~Wallden, ``Optimised resource construction for verifiable
  quantum computation,'' \emph{J. Phys. A: Math. Theor.}, vol.~50, no.~14, p.
  145306, 2017, arXiv:1510.07408, doi: \bibdoi{10.1088/1751-8121/aa5dac}.

\bibitem{gheorghiu_robustness_2015}
A.~Gheorghiu, E.~Kashefi, and P.~Wallden, ``Robustness and device independence
  of verifiable blind quantum computing,'' \emph{New J. Phys.}, vol.~17, no.~8,
  p. 083040, 2015, arXiv:1502.02571, doi:
  \bibdoi{10.1088/1367-2630/17/8/083040}.

\bibitem{leichtle_verifying_2021}
D.~Leichtle, L.~Music, E.~Kashefi, and H.~Ollivier, ``Verifying {BQP}
  {Computations} on {Noisy} {Devices} with {Minimal} {Overhead},'' \emph{PRX
  Quantum}, vol.~2, no.~4, p. 040302, 2021, doi:
  \bibdoi{10.1103/PRXQuantum.2.040302}.

\bibitem{kapourniotis_unifying_2024}
T.~Kapourniotis, E.~Kashefi, D.~Leichtle, L.~Music, and H.~Ollivier, ``Unifying
  quantum verification and error-detection: theory and tools for
  optimisations,'' \emph{Quantum Sci. Technol.}, vol.~9, no.~3, p. 035036,
  2024, arXiv:2206.00631, doi: \bibdoi{10.1088/2058-9565/ad466d}.

\bibitem{inajetovic_verifiable_2025}
M.~Inajetovic, P.~Wallden, and A.~Pappa, ``Verifiable end-to-end delegated
  variational quantum algorithms,'' \emph{Phys. Rev. Research}, vol.~8, no.~3,
  p. 033043, 2026, arXiv:2504.15410, doi: \bibdoi{10.1103/9yqy-8ch9}.

\bibitem{vandam_optimizing_2026}
J.~van Dam, M.~van Hooft, and S.~D.~C. Wehner, ``Optimizing resource costs: A
  practical guide to achieving target security in verifiable blind quantum
  computing,'' arXiv:2606.28139, 2026, doi: \bibdoi{10.48550/arXiv.2606.28139}.

\bibitem{vonneumann_theory_1944}
J.~von Neumann and O.~Morgenstern, \emph{Theory of {Games} and {Economic}
  {Behavior}}.\hskip 1em plus 0.5em minus 0.4em\relax Princeton University
  Press, 1944.

\bibitem{tambe_security_2011}
M.~Tambe, \emph{Security and {Game} {Theory}: {Algorithms}, {Deployed}
  {Systems}, {Lessons} {Learned}}.\hskip 1em plus 0.5em minus 0.4em\relax
  Cambridge University Press, 2011.\hskip 1em plus 0.5em minus 0.4em\relax doi:
  \bibdoi{10.1017/CBO9780511973031}.

\bibitem{korzhyk_stackelberg_2011}
D.~Korzhyk, Z.~Yin, C.~Kiekintveld, V.~Conitzer, and M.~Tambe, ``Stackelberg
  vs. {Nash} in security games: An extended investigation of
  interchangeability, equivalence, and uniqueness,'' \emph{J. Artif. Intell.
  Res.}, vol.~41, pp. 297--327, 2011, doi: \bibdoi{10.1613/jair.3269}.

\bibitem{korzhyk_multiple_2011}
D.~Korzhyk, V.~Conitzer, and R.~Parr, ``Security games with multiple attacker
  resources,'' in \emph{Proceedings of the Twenty-Second International Joint
  Conference on Artificial Intelligence ({IJCAI})}.\hskip 1em plus 0.5em minus
  0.4em\relax AAAI Press/International Joint Conferences on Artificial
  Intelligence, 2011, pp. 273--279, doi:
  \bibdoi{10.5591/978-1-57735-516-8/IJCAI11-056}.

\bibitem{kiekintveld_interval_2013}
\BIBentryALTinterwordspacing
C.~Kiekintveld, T.~Islam, and V.~Kreinovich, ``Security games with interval
  uncertainty,'' in \emph{Proceedings of the 2013 International Conference on
  Autonomous Agents and Multi-Agent Systems ({AAMAS})}.\hskip 1em plus 0.5em
  minus 0.4em\relax International Foundation for Autonomous Agents and
  Multiagent Systems, 2013, pp. 231--238. [Online]. Available:
  \url{https://www.ifaamas.org/Proceedings/aamas2013/docs/p231.pdf}
\BIBentrySTDinterwordspacing

\bibitem{raussendorf_oneway_2001}
R.~Raussendorf and H.~J. Briegel, ``A one-way quantum computer,'' \emph{Phys.
  Rev. Lett.}, vol.~86, no.~22, pp. 5188--5191, 2001, doi:
  \bibdoi{10.1103/PhysRevLett.86.5188}.

\bibitem{briegel_mbqc_2009}
H.~J. Briegel, D.~E. Browne, W.~D{\"u}r, R.~Raussendorf, and M.~Van~den Nest,
  ``Measurement-based quantum computation,'' \emph{Nat. Phys.}, vol.~5, pp.
  19--26, 2009, doi: \bibdoi{10.1038/nphys1157}.

\bibitem{hein_graph_2006}
M.~Hein, W.~D{\"u}r, J.~Eisert, R.~Raussendorf, M.~Van~den Nest, and H.-J.
  Briegel, ``Entanglement in graph states and its applications,'' in
  \emph{Quantum Computers, Algorithms and Chaos}, ser. Proceedings of the
  International School of Physics ``Enrico Fermi''.\hskip 1em plus 0.5em minus
  0.4em\relax IOS Press, 2006, vol. 162, pp. 115--218, arXiv:quant-ph/0602096,
  doi: \bibdoi{10.3254/978-1-61499-018-5-115}.

\bibitem{danos_determinism_2006}
V.~Danos and E.~Kashefi, ``Determinism in the one-way model,'' \emph{Phys. Rev.
  A}, vol.~74, no.~5, p. 052310, 2006, arXiv:quant-ph/0506062, doi:
  \bibdoi{10.1103/PhysRevA.74.052310}.

\bibitem{danos_measurement_2007}
V.~Danos, E.~Kashefi, and P.~Panangaden, ``The measurement calculus,'' \emph{J.
  ACM}, vol.~54, no.~2, p.~8, 2007, arXiv:quant-ph/0412135, doi:
  \bibdoi{10.1145/1219092.1219096}.

\bibitem{browne_generalized_2007}
D.~E. Browne, E.~Kashefi, M.~Mhalla, and S.~Perdrix, ``Generalized flow and
  determinism in measurement-based quantum computation,'' \emph{New J. Phys.},
  vol.~9, no.~8, p. 250, 2007, doi: \bibdoi{10.1088/1367-2630/9/8/250}.

\bibitem{dunjko_composable_2014}
V.~Dunjko, J.~F. Fitzsimons, C.~Portmann, and R.~Renner, ``Composable
  {Security} of {Delegated} {Quantum} {Computation},'' in \emph{Advances in
  {Cryptology} -- {ASIACRYPT} 2014}, ser. Lecture Notes in Computer Science,
  vol. 8874.\hskip 1em plus 0.5em minus 0.4em\relax Springer, 2014, pp.
  406--425, arXiv:1301.3662, doi: \bibdoi{10.1007/978-3-662-45608-8_22}.

\bibitem{morimae_blind_2013}
T.~Morimae and K.~Fujii, ``Blind quantum computation protocol in which {Alice}
  only makes measurements,'' \emph{Phys. Rev. A}, vol.~87, no.~5, p. 050301,
  2013, doi: \bibdoi{10.1103/PhysRevA.87.050301}.

\bibitem{ringbauer_verifiable_2025}
M.~Ringbauer, M.~Hinsche, T.~Feldker, P.~K. Faehrmann, J.~Bermejo-Vega, C.~L.
  Edmunds, L.~Postler, R.~Stricker, C.~D. Marciniak, M.~Meth, I.~Pogorelov,
  R.~Blatt, P.~Schindler, J.~Eisert, T.~Monz, and D.~Hangleiter, ``Verifiable
  measurement-based quantum random sampling with trapped ions,'' \emph{Nat.
  Commun.}, vol.~16, p. 106, 2025, doi: \bibdoi{10.1038/s41467-024-55342-3}.

\bibitem{dankert_exact_2009}
C.~Dankert, R.~Cleve, J.~Emerson, and E.~Livine, ``Exact and approximate
  unitary 2-designs and their application to fidelity estimation,'' \emph{Phys.
  Rev. A}, vol.~80, no.~1, p. 012304, 2009, doi:
  \bibdoi{10.1103/PhysRevA.80.012304}.

\bibitem{nielsen_quantum_2010}
M.~A. Nielsen and I.~L. Chuang, \emph{Quantum {Computation} and {Quantum}
  {Information}}, 10th~ed.\hskip 1em plus 0.5em minus 0.4em\relax Cambridge
  University Press, 2010.\hskip 1em plus 0.5em minus 0.4em\relax doi:
  \bibdoi{10.1017/CBO9780511976667}.

\bibitem{wallman_noise_2016}
J.~J. Wallman and J.~Emerson, ``Noise tailoring for scalable quantum
  computation via randomized compiling,'' \emph{Phys. Rev. A}, vol.~94, no.~5,
  p. 052325, 2016, doi: \bibdoi{10.1103/PhysRevA.94.052325}.

\bibitem{flammia_efficient_2020}
S.~T. Flammia and J.~J. Wallman, ``Efficient {Estimation} of {Pauli}
  {Channels},'' \emph{ACM Trans. Quantum Comput.}, vol.~1, no.~1, pp. 1--32,
  2020, arXiv:1907.12976, doi: \bibdoi{10.1145/3408039}.

\bibitem{harper_efficient_2020}
R.~Harper, S.~T. Flammia, and J.~J. Wallman, ``Efficient learning of quantum
  noise,'' \emph{Nat. Phys.}, vol.~16, no.~12, pp. 1184--1188, 2020, doi:
  \bibdoi{10.1038/s41567-020-0992-8}.

\bibitem{harper_fast_2021}
R.~Harper, W.~Yu, and S.~T. Flammia, ``Fast {Estimation} of {Sparse} {Quantum}
  {Noise},'' \emph{PRX Quantum}, vol.~2, no.~1, p. 010322, 2021, doi:
  \bibdoi{10.1103/PRXQuantum.2.010322}.

\bibitem{chen_learnability_2023}
S.~Chen, Y.~Liu, M.~Otten, A.~Seif, B.~Fefferman, and L.~Jiang, ``The
  learnability of {Pauli} noise,'' \emph{Nat. Commun.}, vol.~14, p.~52, 2023,
  doi: \bibdoi{10.1038/s41467-022-35759-4}.

\bibitem{sion_general_1958}
M.~Sion, ``On general minimax theorems,'' \emph{Pacific J. Math.}, vol.~8,
  no.~1, pp. 171--176, 1958, doi: \bibdoi{10.2140/pjm.1958.8.171}.

\bibitem{nguyen_zerosum_2022}
A.~T. Nguyen, S.~C. Anand, and A.~M.~H. Teixeira, ``A {Zero}-{Sum} {Game}
  {Framework} for {Optimal} {Sensor} {Placement} in {Uncertain} {Networked}
  {Control} {Systems} under {Cyber}-{Attacks},'' in \emph{2022 IEEE 61st
  Conference on Decision and Control (CDC)}.\hskip 1em plus 0.5em minus
  0.4em\relax IEEE, 2022, pp. 6126--6133, doi:
  \bibdoi{10.1109/CDC51059.2022.9992468}.

\bibitem{cover_elements_2006}
T.~M. Cover and J.~A. Thomas, \emph{Elements of {Information} {Theory}},
  2nd~ed.\hskip 1em plus 0.5em minus 0.4em\relax Wiley-Interscience,
  2006.\hskip 1em plus 0.5em minus 0.4em\relax doi:
  \bibdoi{10.1002/047174882X}.

\bibitem{morimae_rational_2020}
T.~Morimae and H.~Nishimura, ``Rational proofs for quantum computing,''
  \emph{Quantum Inf. Comput.}, vol.~20, no. 3--4, pp. 181--193, 2020,
  arXiv:1804.08868, doi: \bibdoi{10.26421/QIC20.3-4-1}.

\bibitem{takeuchi_sumcheck_2019}
Y.~Takeuchi, T.~Morimae, and S.~Tani, ``Sumcheck-{Based} {Delegation} of
  {Quantum} {Computing} to {Rational} {Server},'' in \emph{Theory and
  {Applications} of {Models} of {Computation}}, ser. Lecture Notes in Computer
  Science, vol. 12337.\hskip 1em plus 0.5em minus 0.4em\relax Springer, 2020,
  pp. 69--81, arXiv:1911.04734, doi: \bibdoi{10.1007/978-3-030-59267-7_7}.

\bibitem{meyer_quantum_1999}
D.~A. Meyer, ``Quantum {Strategies},'' \emph{Phys. Rev. Lett.}, vol.~82, no.~5,
  pp. 1052--1055, 1999, doi: \bibdoi{10.1103/PhysRevLett.82.1052}.

\bibitem{eisert_quantum_1999}
J.~Eisert, M.~Wilkens, and M.~Lewenstein, ``Quantum {Games} and {Quantum}
  {Strategies},'' \emph{Phys. Rev. Lett.}, vol.~83, no.~15, pp. 3077--3080,
  1999, arXiv:quant-ph/9806088, doi: \bibdoi{10.1103/PhysRevLett.83.3077}.

\bibitem{hoeffding_probability_1963}
W.~Hoeffding, ``Probability {Inequalities} for {Sums} of {Bounded} {Random}
  {Variables},'' \emph{J. Am. Stat. Assoc.}, vol.~58, no. 301, pp. 13--30,
  1963, doi: \bibdoi{10.1080/01621459.1963.10500830}.

\bibitem{azuma_weighted_1967}
K.~Azuma, ``Weighted sums of certain dependent random variables,'' \emph{Tohoku
  Math. J.}, vol.~19, no.~3, pp. 357--367, 1967, doi:
  \bibdoi{10.2748/tmj/1178243286}.

\bibitem{tsybakov_introduction_2009}
A.~B. Tsybakov, \emph{Introduction to Nonparametric Estimation}, ser. Springer
  Series in Statistics.\hskip 1em plus 0.5em minus 0.4em\relax New York:
  Springer, 2009.\hskip 1em plus 0.5em minus 0.4em\relax doi:
  \bibdoi{10.1007/b13794}.

\bibitem{lattimore_bandit_2020}
T.~Lattimore and C.~Szepesv{\'a}ri, \emph{Bandit Algorithms}.\hskip 1em plus
  0.5em minus 0.4em\relax Cambridge University Press, 2020.\hskip 1em plus
  0.5em minus 0.4em\relax doi: \bibdoi{10.1017/9781108571401}.

\bibitem{flammia_direct_2011}
S.~T. Flammia and Y.-K. Liu, ``Direct {Fidelity} {Estimation} from {Few}
  {Pauli} {Measurements},'' \emph{Phys. Rev. Lett.}, vol. 106, no.~23, p.
  230501, 2011, doi: \bibdoi{10.1103/PhysRevLett.106.230501}.

\bibitem{eisert_certification_2020}
J.~Eisert, D.~Hangleiter, N.~Walk, I.~Roth, D.~Markham, R.~Parekh, U.~Chabaud,
  and E.~Kashefi, ``Quantum certification and benchmarking,'' \emph{Nat. Rev.
  Phys.}, vol.~2, pp. 382--390, 2020, doi: \bibdoi{10.1038/s42254-020-0186-4}.

\bibitem{scheinerman_fractional_1997}
E.~R. Scheinerman and D.~H. Ullman, \emph{Fractional Graph Theory: A Rational
  Approach to the Theory of Graphs}.\hskip 1em plus 0.5em minus 0.4em\relax
  Wiley, 1997.

\bibitem{grotschel_geometric_1988}
M.~Gr{\"o}tschel, L.~Lov{\'a}sz, and A.~Schrijver, \emph{Geometric Algorithms
  and Combinatorial Optimization}, ser. Algorithms and Combinatorics.\hskip 1em
  plus 0.5em minus 0.4em\relax Springer, 1988, vol.~2, doi:
  \bibdoi{10.1007/978-3-642-97881-4}.

\bibitem{schrijver_combinatorial_2003}
A.~Schrijver, \emph{Combinatorial Optimization: Polyhedra and Efficiency}, ser.
  Algorithms and Combinatorics.\hskip 1em plus 0.5em minus 0.4em\relax
  Springer, 2003, vol.~24.

\bibitem{garey_computers_1979}
M.~R. Garey and D.~S. Johnson, \emph{Computers and Intractability: A Guide to
  the Theory of {NP}-Completeness}.\hskip 1em plus 0.5em minus 0.4em\relax W.
  H. Freeman, 1979.

\bibitem{shannon_communication_1949}
C.~E. Shannon, ``Communication {Theory} of {Secrecy} {Systems},'' \emph{Bell
  Syst. Tech. J.}, vol.~28, no.~4, pp. 656--715, 1949, doi:
  \bibdoi{10.1002/j.1538-7305.1949.tb00928.x}.

\end{thebibliography}

\end{document}